\documentclass[english]{article}
\usepackage[T1]{fontenc}
\usepackage[latin9]{inputenc}
\usepackage{mathrsfs}
\usepackage{amsmath}
\usepackage{amssymb}
\usepackage{esint}

\makeatletter
\usepackage{fullpage}

\makeatother

\usepackage{babel}
\begin{document}

\title{Weyl geometry}

\author{James T. Wheeler\thanks{Utah State University Dept of Physics email: jim.wheeler@usu.edu}}
\maketitle
\begin{abstract}
We develop the properties of Weyl geometry, beginning with a review
of the conformal properties of Riemannian spacetimes. Decomposition
of the Riemann curvature into trace and traceless parts allows an
easy proof that the Weyl curvature tensor is the conformally invariant
part of the Riemann curvature, and shows the explicit change in the
Ricci and Schouten tensors required to insure conformal invariance.
We include a proof of the well-known condition for the existence of
a conformal transformation to a Ricci-flat spacetime. We generalize
this to a derivation of the condition for the existence of a conformal
transformation to a spacetime satisfying the Einstein equation with
matter sources. Then, enlarging the symmetry from Poincaré to Weyl,
we develop the Cartan structure equations of Weyl geometry, the form
of the curvature tensor and its relationship to the Riemann curvature
of the corresponding Riemannian geometry. We present a simple theory
of Weyl-covariant gravity based on a curvature-linear action, and
show that it is conformally equivalent to general relativity. This
theory is invariant under local dilatations, but not the full conformal
group.
\end{abstract}

\section{Introduction}

In 1918, H. Weyl introduced an additional symmetry into Riemannian
geometry in an attempt to unify electromagnetism with gravity as a
fully geometric model \cite{Weyl 1918a,Weyl 1918b}. The idea was
to allow both the orientation and the length of vectors to vary under
parallel transport, instead of just the orientation as in Riemannian
geometry. The resulting \emph{geometries} are called Weyl geometries,
and they form a completely consistent generalization of Riemannian
geometries. However, Weyl's attempt to identify the vector part of
the connection associated with stretching and contraction with the
vector potential of electromagnetism failed. As Einstein pointed out
immediately following Weyl's first paper on the subject \cite{Einstein},
the identification implies that identical atoms which move in such
a way as to enclose some electromagnetic flux would be different sizes
after the motion. Different sized atoms would have different spectra,
and it is easy to show that change in frequency resulting from the
size change would be vastly inconsistent with the known precision
of spectral lines.

Many attempts were made to patch up the theory. In the end, following
some notable work by London \cite{London}, Weyl showed that a satisfactory
theory of electromagnetism is achieved if the scale factor is replaced
by a complex phase. This is the origin of $U\left(1\right)$ gauge
theory. Many interesting details are discussed in O'Raifeartaigh \cite{O'Raifeartaigh},
and a basic introduction to Weyl geometry is given in \cite{AdlerBazinSchiffer}.

In modern language, the new vector part of the connection introduced
by Weyl is the dilatational gauge vector, often called the Weyl vector.
When this vector is given by the gradient of a function, then there
exists a scale transformation (understandable as a change of units,
or a dilatation) that sets the vector to zero. When this is possible,
the Weyl geometry is called \emph{integrable}: vectors parallel transported
about closed paths return with their lengths unaltered. Integrable
Weyl geometries are trivial in the sense that there exists a subclass
of global gauges in which the geometry is Riemannian.

While Weyl's theory of electromagnetism fails, Weyl geometry does
not. Indeed, although no new physical predictions have emerged directly
from its use, there are at least the following three considerations
for seeking a deeper understanding of general relativity formulated
within an integrable Weyl geometry:
\begin{enumerate}
\item General relativity is naturally invariant under global changes of
units. By formulating general relativity in an integrable Weyl geometry,
this scale invariance becomes local. We refer to this generalization
as \emph{scale invariant general relativity.} As soon as we make a
definition of a fundamental standard of length \textendash{} for example,
as the distance light travels in one second\footnote{The second is currently defined as the duration of $9,192,631,770$
periods of the radiation corresponding to the transition between the
two hyperfine levels of the ground state of the caesium 133 atom {[}physics.nist.gov{]}.
The metre is defined as the length of the path travelled by light
in a vacuum in $\frac{1}{299792458}$ second {[}17th General Conference
on Weights and Measures (1983), Resolution 1{]}.} \textendash{} scale invariant general relativity reduces to general
relativity.
\item In \cite{EhlersPiraniSchild}, Ehlers, Pirani and Schild make the
following argument. First, the paths of light pulses may be used to
determine a conformal connection on spacetime. Then, a projective
connection is found by tracing trajectories of massive test particles
(``dust''). Finally, requiring the two connections to approach one
another in the limit of near-light velocities reduces the possible
connection to that of a Weyl geometry. When this program is carried
out with mathematical precision \cite{MatveevTrautman}, the resulting
geometry is an \emph{integrable Weyl geometry}.
\item Deeper physical interest in Weyl geometry also arises in higher symmetry
approaches to gravity. Gravitational theories based on the full conformal
group (\cite{Weyl 1918a},\cite{Bach}-\cite{WW}) often yield general
relativity formulated on an integrable Weyl geometry rather than on
a Riemannian one and are therefore equivalent to general relativity
while providing additional natural structures.
\end{enumerate}
For these reasons, it is useful to recognize the typical forms and
meaning of the connection and curvature of Weyl geometry.

Here we use the techniques of modern gauge theory \cite{Kobayashi and Nomizu,Neeman and Regge}
to develop the properties of Weyl geometry, beginning in the next
section with a review of the conformal properties of Riemannian spacetimes.
Decomposition of the Riemann curvature into trace and traceless parts
allows an easy proof that the Weyl curvature tensor is the conformally
invariant part of the Riemann curvature, and shows the explicit change
in the Ricci and Schouten tensors required to insure conformal invariance.
We include a proof of the well-known condition for the existence of
a conformal transformation to a Ricci-flat spacetime, and generalize
this to a derivation of the condition for the existence of a conformal
transformation to a spacetime satisfying the Einstein equation with
matter sources. Then, in the final section, we enlarge the symmetry
from Poincaré to Weyl to develop the Cartan structure equations of
Weyl geometry, the form of the curvature tensor, and its relationship
to the Riemann curvature of the corresponding Riemannian geometry.
We conclude with a simple theory of Weyl-covariant gravity based on
a curvature-linear action, and show that its vacuum solutions are
conformal equivalence classes of Ricci-flat metrics in an integrable
Weyl geometry. This theory is invariant under dilatations, but not
the full conformal group.

\section{Conformal transformations in Riemannian geometry\label{Conformal transforms in Riemannian geometry}}

\subsection{Structure equations for Riemannian geometry}

The Cartan structure equations of a Riemannian geometry are
\begin{eqnarray}
\mathbf{R}_{\;b}^{a} & = & \mathbf{d}\boldsymbol{\alpha}_{\;b}^{a}-\boldsymbol{\alpha}_{\;b}^{c}\wedge\boldsymbol{\alpha}_{\;c}^{a}\label{Structure equation for spin connection}\\
0 & = & \mathbf{d}\mathbf{e}^{a}-\mathbf{e}^{b}\wedge\boldsymbol{\alpha}_{\;b}^{a}\label{Structure equation for solder form}
\end{eqnarray}
where the solder form, $\mathbf{e}^{a}=e_{\mu}^{\quad a}\mathbf{d}x^{\mu}$,
provides an orthonormal basis, $\boldsymbol{\alpha}_{\;b}^{a}$ is
the spin connection $1$-form, and the curvature $2$-form is $\mathbf{R}_{\;b}^{a}=\frac{1}{2}R_{\;bcd}^{a}\mathbf{e}^{c}\wedge\mathbf{e}^{d}$.
Differential forms are written in boldface.

The structure equations satisfy integrability conditions, the Bianchi
identities, found by applying the Poincaré lemma, $\mathbf{d}^{2}\equiv0$:
\begin{eqnarray*}
0\equiv\mathbf{d}^{2}\boldsymbol{\alpha}_{\;b}^{a} & = & \mathbf{d}\left(\boldsymbol{\alpha}_{\;b}^{c}\wedge\boldsymbol{\alpha}_{\;c}^{a}+\mathbf{R}_{\;b}^{a}\right)\\
 & = & \mathbf{d}\boldsymbol{\alpha}_{\;b}^{c}\wedge\boldsymbol{\alpha}_{\;c}^{a}-\boldsymbol{\alpha}_{\;b}^{c}\wedge\mathbf{d}\boldsymbol{\alpha}_{\;c}^{a}+\mathbf{d}\mathbf{R}_{\;b}^{a}\\
 & = & \left(\boldsymbol{\alpha}_{\;b}^{e}\wedge\boldsymbol{\alpha}_{\;e}^{c}+\mathbf{R}_{\;b}^{c}\right)\wedge\boldsymbol{\alpha}_{\;c}^{a}-\boldsymbol{\alpha}_{\;b}^{c}\wedge\left(\boldsymbol{\alpha}_{\;c}^{e}\wedge\boldsymbol{\alpha}_{\;e}^{a}+\mathbf{R}_{\;c}^{a}\right)+\mathbf{d}\mathbf{R}_{\;b}^{a}\\
 & = & \mathbf{d}\mathbf{R}_{\;b}^{a}+\mathbf{R}_{\;b}^{c}\wedge\boldsymbol{\alpha}_{\;c}^{a}-\boldsymbol{\alpha}_{\;b}^{c}\wedge\mathbf{R}_{\;c}^{a}\\
 & \equiv & \mathbf{D}\mathbf{R}_{\;b}^{a}\\
0\equiv\mathbf{d}^{2}\mathbf{e}^{a} & = & \mathbf{d}\left(\mathbf{e}^{b}\wedge\boldsymbol{\alpha}_{\;b}^{a}\right)\\
 & = & \mathbf{d}\mathbf{e}^{b}\wedge\boldsymbol{\alpha}_{\;b}^{a}-\mathbf{e}^{b}\wedge\mathbf{d}\boldsymbol{\alpha}_{\;b}^{a}\\
 & = & \left(\mathbf{e}^{c}\wedge\boldsymbol{\alpha}_{\;c}^{c}\right)\wedge\boldsymbol{\alpha}_{\;b}^{a}-\mathbf{e}^{b}\wedge\left(\boldsymbol{\alpha}_{\;b}^{c}\wedge\boldsymbol{\alpha}_{\;c}^{a}+\mathbf{R}_{\;b}^{a}\right)\\
 & = & -\mathbf{e}^{b}\wedge\mathbf{R}_{\;b}^{a}
\end{eqnarray*}
In components, these take the familiar forms
\begin{eqnarray*}
R_{\;b\left[cd;e\right]}^{a} & = & 0\\
R_{\;\left[bcd\right]}^{a} & = & 0
\end{eqnarray*}
Here we use Greek and Latin indices to distinguish different vector
bases. Use of the covariantly constant coefficient matrix of the solder
form, $e_{\mu}^{\quad a}$, allows us to convert freely between orthonormal
components (Latin indices) and coordinate components (Greek indices),
$R_{\;\beta\mu\nu}^{\alpha}=e_{a}^{\quad\alpha}e_{\beta}^{\quad b}e_{\mu}^{\quad c}e_{\nu}^{\quad d}R_{\;bcd}^{a}$.

\subsection{Conformal transformation of the metric, solder form, and connection }

A conformal transformation of the metric is the transformation
\begin{equation}
g_{\mu\nu}\rightarrow\tilde{g}_{\mu\nu}=e^{2\phi}g_{\mu\nu}\label{Conformal transformation}
\end{equation}
This is not an invariance of Riemannian geometry, but it is an invariance
of Weyl geometry. If we make a change of this type in a Riemannian
geometry, the solder form changes by
\begin{equation}
\mathbf{e}^{a}\rightarrow\tilde{\mathbf{e}}^{a}=e^{\phi}\mathbf{e}^{a}\label{Transformation of the solder form}
\end{equation}
since the solder form and metric are related via the orthonormal metric,
$\eta_{ab}=diag\left(-1,1,1,1\right)$ by
\[
g_{\mu\nu}=e_{\mu}^{\quad a}e_{\nu}^{\quad b}\eta_{ab}
\]
The corresponding structure equation then gives the altered form of
the metric compatible connection,
\begin{eqnarray*}
\mathbf{d}\tilde{\mathbf{e}}^{a} & = & \tilde{\mathbf{e}}^{b}\wedge\boldsymbol{\tilde{\alpha}}_{\;b}^{a}\\
\mathbf{d}\left(e^{\phi}\mathbf{e}^{a}\right) & = & e^{\phi}\mathbf{e}^{b}\wedge\boldsymbol{\tilde{\alpha}}_{\;b}^{a}\\
e^{\phi}\left(\mathbf{d}\mathbf{e}^{a}+\mathbf{d}\phi\wedge\mathbf{e}^{a}\right) & = & e^{\phi}\mathbf{e}^{b}\wedge\boldsymbol{\tilde{\alpha}}_{\;b}^{a}
\end{eqnarray*}
so to find the new spin connection we must solve
\begin{equation}
\mathbf{d}\mathbf{e}^{a}=\mathbf{e}^{b}\wedge\boldsymbol{\tilde{\alpha}}_{\;b}^{a}-\mathbf{d}\phi\wedge\mathbf{e}^{a}\label{Structure equation for conformally transformed spacetime}
\end{equation}
Since the spin connection is antisymmetric, $\boldsymbol{\tilde{\alpha}}_{\;b}^{a}=-\eta^{ad}\eta_{bc}\boldsymbol{\tilde{\alpha}}_{\;d}^{c}$,
this is solved by setting
\begin{equation}
\boldsymbol{\tilde{\alpha}}_{\;b}^{a}=\boldsymbol{\alpha}_{\;b}^{a}+2\Delta_{db}^{ac}e_{c}^{\quad\mu}\partial_{\mu}\phi\,\mathbf{e}^{d}\label{Transformation of the spin connection}
\end{equation}
where $e_{c}^{\quad\mu}$ is inverse to $e_{\mu}^{\quad a}$. The
convenient symbol $\Delta_{db}^{ac}$ is defined as 
\begin{equation}
\Delta_{db}^{ac}\equiv\frac{1}{2}\left(\delta_{d}^{a}\delta_{b}^{c}-\eta^{ac}\eta_{bd}\right)\label{Definition of Delta}
\end{equation}
This is meant to act on any $\left(\begin{array}{c}
1\\
1
\end{array}\right)$ tensor according to
\begin{eqnarray*}
\Delta_{db}^{ac}T_{\;\;c}^{d} & \equiv & \frac{1}{2}\left(\delta_{d}^{a}\delta_{b}^{c}-\eta^{ac}\eta_{bd}\right)T_{\;\;c}^{d}\\
 & = & \frac{1}{2}\left(T_{\;\;b}^{a}-\eta^{ac}\eta_{bd}T_{\;\;c}^{d}\right)\\
 & = & \frac{1}{2}\eta^{ac}\left(T_{cb}-T_{bc}\right)
\end{eqnarray*}
which is just an antisymmetric $\left(\begin{array}{c}
0\\
2
\end{array}\right)$ tensor with the first index raised to give a $\left(\begin{array}{c}
1\\
1
\end{array}\right)$ tensor. It is a projection since it is idempotent,
\begin{eqnarray*}
\Delta_{db}^{ac}\Delta_{cf}^{ed} & = & \frac{1}{4}\left(\delta_{d}^{a}\delta_{b}^{c}-\eta^{ac}\eta_{bd}\right)\left(\delta_{c}^{e}\delta_{f}^{d}-\eta^{ed}\eta_{cf}\right)\\
 & = & \frac{1}{4}\left(\delta_{b}^{e}\delta_{f}^{a}-\eta^{ae}\eta_{bf}-\eta^{ea}\eta_{bf}+\delta_{f}^{a}\delta_{b}^{e}\right)\\
 & = & \Delta_{fb}^{ae}
\end{eqnarray*}

Returning to check eq.(\ref{Transformation of the spin connection})
in eq.(\ref{Structure equation for conformally transformed spacetime}),
we have,
\begin{eqnarray*}
\mathbf{d}\mathbf{e}^{a} & = & \mathbf{e}^{b}\wedge\boldsymbol{\tilde{\alpha}}_{\;b}^{a}-\mathbf{d}\phi\wedge\mathbf{e}^{a}\\
 & = & \mathbf{e}^{b}\wedge\left(\boldsymbol{\alpha}_{\;b}^{a}+2\Delta_{db}^{ac}e_{c}^{\quad\mu}\partial_{\mu}\phi\,\mathbf{e}^{d}\right)-\mathbf{d}\phi\wedge\mathbf{e}^{a}\\
 & = & \mathbf{e}^{b}\wedge\boldsymbol{\alpha}_{\;b}^{a}+\left(\delta_{d}^{a}\delta_{b}^{c}-\eta^{ac}\eta_{bd}\right)e_{c}^{\quad\mu}\partial_{\mu}\phi\,\mathbf{e}^{b}\wedge\mathbf{e}^{d}-\mathbf{d}\phi\wedge\mathbf{e}^{a}\\
 & = & \mathbf{e}^{b}\wedge\boldsymbol{\alpha}_{\;b}^{a}+e_{b}^{\quad\mu}\partial_{\mu}\phi\,\mathbf{e}^{b}\wedge\mathbf{e}^{a}-\mathbf{d}\phi\wedge\mathbf{e}^{a}\\
 & = & \mathbf{e}^{b}\wedge\boldsymbol{\alpha}_{\;b}^{a}+\mathbf{d}\phi\wedge\mathbf{e}^{a}-\mathbf{d}\phi\wedge\mathbf{e}^{a}\\
 & = & \mathbf{e}^{b}\wedge\boldsymbol{\alpha}_{\;b}^{a}
\end{eqnarray*}
as required. Since the spin connection is uniquely determined (up
to local Lorentz transformations) by the structure equation, this
is the unique solution.

\subsection{Transformation of the curvature}

Now compute the new curvature tensor. For this longer calculation
it is convenient to define the orthonormal component of $\mathbf{d}\phi=\phi_{b}\mathbf{e}^{b}$,
\[
\phi_{a}\equiv e_{a}^{\quad\mu}\partial_{\mu}\phi.
\]
 Then $\boldsymbol{\tilde{\alpha}}_{\;b}^{a}=\boldsymbol{\alpha}_{\;b}^{a}+2\Delta_{db}^{ac}\phi_{c}\,\mathbf{e}^{d}$
and the conformal transformation changes the curvature to:

\begin{eqnarray*}
\frac{1}{2}\tilde{R}_{\;bcd}^{a}\mathbf{e}^{c}\wedge\mathbf{e}^{d} & = & \mathbf{d}\tilde{\boldsymbol{\alpha}}_{\;b}^{a}-\tilde{\boldsymbol{\alpha}}_{\;b}^{c}\wedge\tilde{\boldsymbol{\alpha}}_{\;c}^{a}\\
 & = & \mathbf{d}\boldsymbol{\alpha}_{\;b}^{a}-\boldsymbol{\alpha}_{\;b}^{c}\wedge\boldsymbol{\alpha}_{\;c}^{a}+\mathbf{d}\left(2\Delta_{db}^{ac}\phi_{c}\mathbf{e}^{d}\right)-\boldsymbol{\alpha}_{\;b}^{c}\wedge\left(2\Delta_{dc}^{ae}\phi_{e}\mathbf{e}^{d}\right)\\
 &  & -\left(2\Delta_{db}^{ce}\phi_{e}\mathbf{e}^{d}\right)\wedge\boldsymbol{\alpha}_{\;c}^{a}-\left(2\Delta_{db}^{ce}\phi_{e}\mathbf{e}^{d}\right)\wedge\left(2\Delta_{fc}^{ag}\phi_{g}\mathbf{e}^{f}\right)\\
 & = & \frac{1}{2}R_{\;bcd}^{a}\mathbf{e}^{c}\wedge\mathbf{e}^{d}+\mathbf{D}\left(2\Delta_{db}^{ac}\phi_{c}\mathbf{e}^{d}\right)-\left(2\Delta_{db}^{ce}\phi_{e}\mathbf{e}^{d}\right)\wedge\left(2\Delta_{fc}^{ag}\phi_{g}\mathbf{e}^{f}\right)
\end{eqnarray*}
Since 
\[
\mathbf{D}\mathbf{e}^{a}=\mathbf{d}\mathbf{e}^{a}-\mathbf{e}^{c}\wedge\boldsymbol{\alpha}_{\;c}^{a}=0
\]
this reduces to
\[
\frac{1}{2}\tilde{R}_{\;bcd}^{a}\mathbf{e}^{c}\wedge\mathbf{e}^{d}=\frac{1}{2}R_{\;bcd}^{a}\mathbf{e}^{c}\wedge\mathbf{e}^{d}+2\Delta_{db}^{ac}\mathbf{D}\phi_{c}\mathbf{e}^{d}-\left(2\Delta_{db}^{ce}\phi_{e}\mathbf{e}^{d}\right)\wedge\left(2\Delta_{fc}^{ag}\phi_{g}\mathbf{e}^{f}\right)
\]
as is easily checked by expanding the $\Delta s$. We need only simplify
the final term:
\begin{eqnarray*}
4\Delta_{db}^{ce}\phi_{e}\mathbf{e}^{d}\wedge\Delta_{fc}^{ag}\phi_{g}\mathbf{e}^{f} & = & \left(2\Delta_{db}^{ce}\phi_{e}\mathbf{e}^{d}\right)\wedge\left(2\Delta_{fc}^{ag}\phi_{g}\mathbf{e}^{f}\right)\\
 & = & \left(\phi_{b}\mathbf{e}^{c}-\eta^{ce}\eta_{bd}\phi_{e}\mathbf{e}^{d}\right)\wedge\left(\phi_{c}\mathbf{e}^{a}-\eta^{ag}\eta_{fc}\phi_{g}\mathbf{e}^{f}\right)\\
 & = & \phi_{c}\phi_{b}\mathbf{e}^{c}\wedge\mathbf{e}^{a}-\eta^{ag}\phi_{g}\phi_{b}\eta_{fc}\mathbf{e}^{c}\wedge\mathbf{e}^{f}-\eta_{bd}\eta^{ce}\phi_{c}\phi_{e}\mathbf{e}^{d}\wedge\mathbf{e}^{a}\\
 &  & +\eta^{ag}\eta_{fc}\phi_{g}\eta^{ce}\eta_{bd}\phi_{e}\mathbf{e}^{d}\wedge\mathbf{e}^{f}\\
 & = & \left(\phi_{b}\phi_{f}\mathbf{e}^{f}\wedge\mathbf{e}^{a}-\eta_{bd}\phi^{c}\phi_{c}\mathbf{e}^{d}\wedge\mathbf{e}^{a}-\eta^{ag}\eta_{bd}\phi_{g}\phi_{f}\mathbf{e}^{f}\wedge\mathbf{e}^{d}\right)\\
 & = & \left(\left(\delta_{d}^{a}\delta_{b}^{e}\phi_{e}\phi_{f}-\eta^{ae}\eta_{bd}\phi_{e}\phi_{f}\right)\mathbf{e}^{f}\wedge\mathbf{e}^{d}-\eta_{bd}\phi^{c}\phi_{c}\mathbf{e}^{d}\wedge\mathbf{e}^{a}\right)\\
 & = & \left(\left(\delta_{d}^{a}\delta_{b}^{c}-\eta^{ac}\eta_{bd}\right)\phi_{c}\phi_{e}\mathbf{e}^{e}\wedge\mathbf{e}^{d}-\frac{1}{2}\delta_{d}^{a}\eta_{bc}\phi^{f}\phi_{f}\mathbf{e}^{c}\wedge\mathbf{e}^{d}+\frac{1}{2}\delta_{c}^{a}\eta_{bd}\phi^{f}\phi_{f}\mathbf{e}^{c}\wedge\mathbf{e}^{d}\right)\\
 & = & \left(2\Delta_{db}^{ac}\phi_{c}\phi_{e}-\frac{1}{2}\delta_{d}^{a}\delta_{b}^{c}\eta_{ce}\phi^{f}\phi_{f}+\frac{1}{2}\eta^{ac}\eta_{bd}\eta_{ce}\phi^{f}\phi_{f}\right)\mathbf{e}^{e}\wedge\mathbf{e}^{d}\\
 & = & 2\Delta_{db}^{ac}\left(\phi_{c}\phi_{e}-\frac{1}{2}\eta_{ce}\phi^{f}\phi_{f}\right)\mathbf{e}^{e}\wedge\mathbf{e}^{d}
\end{eqnarray*}
Now setting $\mathbf{D}^{\left(\alpha\right)}\phi_{c}=\phi_{c;d}\mathbf{e}^{d}$
and $\mathbf{d}^{\left(x\right)}\phi=\phi_{d}\mathbf{e}^{d}$. the
final result is, 
\begin{equation}
\tilde{\mathbf{R}}_{\;b}^{a}=\mathbf{R}_{\;b}^{a}+2\Delta_{db}^{ac}\left(\phi_{c;e}-\phi_{e}\phi_{c}+\frac{1}{2}\phi^{f}\phi_{f}\eta_{ce}\right)\mathbf{e}^{e}\wedge\mathbf{e}^{d}\label{Change in curvature}
\end{equation}
In components, eq.(\ref{Change in curvature}) becomes
\[
\tilde{R}_{\;bcd}^{a}=R_{\;bcd}^{a}+2\Delta_{db}^{ae}\left(\phi_{e;c}-\phi_{c}\phi_{e}+\frac{1}{2}\phi^{f}\phi_{f}\eta_{ec}\right)-2\Delta_{cb}^{ae}\left(\phi_{e;d}-\phi_{d}\phi_{e}+\frac{1}{2}\phi^{f}\phi_{f}\eta_{ed}\right)
\]
The Ricci tensor and scalar are 
\begin{eqnarray*}
\tilde{R}_{bd} & = & R_{bd}-\left(n-2\right)\phi_{b;d}-\eta_{bd}\phi_{\:;c}^{c}+\left(n-2\right)\phi_{b}\phi_{d}+\left(n-2\right)\eta_{bd}\phi^{c}\phi_{c}
\end{eqnarray*}
and
\begin{eqnarray*}
\tilde{R} & = & \tilde{g}^{ab}\tilde{R}_{ab}\\
 & = & e^{-2\phi}g^{ab}\left(R_{bd}-\left(n-2\right)\phi_{b;d}-\eta^{ce}\phi_{e;c}\eta_{bd}+\left(n-2\right)\phi_{d}\phi_{b}-\left(n-2\right)\phi^{c}\phi_{c}\eta_{bd}\right)\\
 & = & e^{-2\phi}\left(R-2\left(n-1\right)\phi_{\;\;;c}^{c}+\left(n-2\right)\phi^{c}\phi_{c}-n\left(n-2\right)\phi^{c}\phi_{c}\right)\\
 & = & e^{-2\phi}\left(R-2\left(n-1\right)\phi_{\;\;;c}^{c}-\left(n-1\right)\left(n-2\right)\phi^{c}\phi_{c}\right)
\end{eqnarray*}

\subsection{Invariance of the Weyl curvature tensor}

In general, we may split the Riemann curvature $R_{\;bcd}^{a}$ into
its trace, the Ricci tensor, and its traceless part, called the Weyl
curvature. This decomposition is most concisely expressed if we first
define the Schouten tensor,
\begin{equation}
\mathcal{R}_{bd}\equiv\frac{1}{\left(n-2\right)}\left(R_{bd}-\frac{1}{2\left(n-1\right)}\eta_{db}R\right)\label{Schouten tensor}
\end{equation}
where $R_{ab}$ is the Ricci tensor,
\[
R_{ab}\equiv R_{\;acb}^{c}
\]
The Schouten tensor often arises as a $1$-form, $\boldsymbol{\mathcal{R}}_{a}=\mathcal{R}_{ab}\mathbf{e}^{b}$.
Except in $2$-dimensions, it is equivalent to the Ricci tensor, since
we may invert eq.(\ref{Schouten tensor}) to write 
\begin{eqnarray}
R_{bd} & = & \left(n-2\right)\mathcal{R}_{bd}+\eta_{db}\mathcal{R}\label{Ricci in terms of Schouten}\\
R & = & 2\left(n-1\right)\mathcal{R}\label{Ricci scalar in terms of Schouten}
\end{eqnarray}

In terms of $\mathcal{R}_{ab}$, the Weyl curvature $2$-form is defined
as
\begin{equation}
\mathbf{C}_{\;b}^{a}\equiv\mathbf{R}_{\;b}^{a}+2\Delta_{db}^{ae}\boldsymbol{\mathcal{R}}_{e}\wedge\mathbf{e}^{d}\label{Weyl curvature}
\end{equation}
Expanding to find the components, $C_{\;bcd}^{a}$, of $\mathbf{C}_{\;b}^{a}$,
\begin{eqnarray}
C_{\;bcd}^{a} & = & R_{\;bcd}^{a}+2\Delta_{db}^{ae}\mathcal{R}_{ec}-2\Delta_{cb}^{ae}\mathcal{R}_{ed}\nonumber \\
 & = & R_{\;bcd}^{a}+\left(\delta_{d}^{a}\delta_{b}^{e}-\eta^{ae}\eta_{bd}\right)\mathcal{R}_{ec}-\left(\delta_{c}^{a}\delta_{b}^{e}-\eta^{ae}\eta_{cb}\right)\mathcal{R}_{ed}\nonumber \\
 & = & R_{\;bcd}^{a}+\delta_{d}^{a}\mathcal{R}_{bc}-\mathcal{R}_{\:c}^{a}\eta_{bd}-\delta_{c}^{a}\mathcal{R}_{bd}+\mathcal{R}_{\:d}^{a}\eta_{bc}\nonumber \\
 & = & R_{\;bcd}^{a}-\frac{1}{\left(n-2\right)}\left(\delta_{c}^{a}R_{bd}-\delta_{d}^{a}R_{bc}-R_{\:d}^{a}\eta_{bc}+R_{\:c}^{a}\eta_{bd}\right)+\frac{R}{\left(n-1\right)\left(n-2\right)}\left(\delta_{c}^{a}\eta_{bd}-\delta_{d}^{a}\eta_{bc}\right)\label{Weyl curvature in components}
\end{eqnarray}
we readily verify its tracelessness,
\begin{eqnarray*}
C_{\;bcd}^{c} & = & R_{bd}-\frac{1}{n-2}\left(nR_{bd}-R_{bd}-R_{bd}+\eta_{bd}R\right)+\frac{R}{\left(n-1\right)\left(n-2\right)}\left(n-1\right)\eta_{bd}\\
 & = & 0
\end{eqnarray*}
with all other nontrivial traces equivalent to this one. By contrast,
the second term in eq.(\ref{Weyl curvature}) is equivalent to knowing
the Ricci or Schouten tensor, since the components of the $\Delta_{cb}^{ae}$
term, $D_{\:bcd}^{a}\equiv\Delta_{db}^{ae}\mathcal{R}_{ec}-\Delta_{cb}^{ae}\mathcal{R}_{ed}$
in eq.(\ref{Weyl curvature}) give
\begin{eqnarray*}
\mathcal{R}_{bd} & = & \delta_{a}^{c}\left[-\frac{2}{n-2}D_{\:bcd}^{a}+\frac{1}{\left(n-1\right)\left(n-2\right)}\left(\eta^{fg}D_{\:fcg}^{a}\right)\eta_{bd}\right]
\end{eqnarray*}
To check this we expand,
\begin{eqnarray*}
\mathcal{R}_{bd} & = & \delta_{a}^{c}\left[-\frac{2}{n-2}D_{\:bcd}^{a}+\frac{1}{\left(n-1\right)\left(n-2\right)}\left(\eta^{fg}D_{\:fcg}^{a}\right)\eta_{bd}\right]\\
 & = & \delta_{a}^{c}\left[-\frac{2}{n-2}\left(\Delta_{db}^{ae}\mathcal{R}_{ec}-\Delta_{cb}^{ae}\mathcal{R}_{ed}\right)+\frac{1}{\left(n-1\right)\left(n-2\right)}\left(\eta^{fg}\Delta_{fg}^{ae}\mathcal{R}_{ec}-\eta^{fg}\Delta_{cf}^{ae}\mathcal{R}_{eg}\right)\eta_{bd}\right]\\
 & = & -\frac{2}{n-2}\left(\frac{1}{2}\left(\delta_{d}^{a}\delta_{b}^{e}-\eta^{ae}\eta_{db}\right)\mathcal{R}_{ea}-\frac{1}{2}\left(n-1\right)\mathcal{R}_{bd}\right)\\
 &  & +\frac{1}{\left(n-1\right)\left(n-2\right)}\left(\frac{1}{2}\eta^{fg}\left(\delta_{f}^{a}\delta_{g}^{e}-\eta^{ae}\eta_{fg}\right)\mathcal{R}_{ea}-\frac{1}{2}\left(n-1\right)\mathcal{R}\right)\eta_{bd}\\
 & = & -\frac{1}{n-2}\left(\mathcal{R}_{bd}-\eta_{bd}\mathcal{R}-\left(n-1\right)\mathcal{R}_{bd}\right)+\frac{1}{2\left(n-1\right)\left(n-2\right)}\left(1-n-\left(n-1\right)\right)\mathcal{R}\eta_{bd}\\
 & = & \frac{1}{n-2}\left[-\mathcal{R}_{bd}+\left(n-1\right)\mathcal{R}_{bd}+\eta_{bd}\mathcal{R}-\mathcal{R}\eta_{bd}\right]\\
 & = & \mathcal{R}_{bd}
\end{eqnarray*}

We now have the decomposition of the Riemann curvature into traceless
and trace parts,
\begin{equation}
\mathbf{R}_{\;b}^{a}=\mathbf{C}_{\;b}^{a}-2\Delta_{db}^{ae}\boldsymbol{\mathcal{R}}_{e}\wedge\mathbf{e}^{d}\label{Decomposition of the Riemann curvature}
\end{equation}
After a conformal transformation, the new Riemann curvature $2$-form
may also be decomposed in the same way,
\[
\tilde{\mathbf{R}}_{\;b}^{a}=\tilde{\mathbf{C}}_{\;b}^{a}-2\Delta_{db}^{ae}\tilde{\boldsymbol{\mathcal{R}}}_{e}\wedge\tilde{\mathbf{e}}^{d}
\]
Combining this with eq.(\ref{Change in curvature}) we have
\begin{eqnarray*}
\tilde{\mathbf{C}}_{\;b}^{a}-2\Delta_{db}^{ae}\tilde{\boldsymbol{\mathcal{R}}}_{e}\wedge\tilde{\mathbf{e}}^{d} & = & \mathbf{C}_{\;b}^{a}-2\Delta_{db}^{ae}\boldsymbol{\mathcal{R}}_{e}\wedge\mathbf{e}^{d}+2\Delta_{db}^{ae}\left(\phi_{e;c}-\phi_{e}\phi_{c}+\frac{1}{2}\phi^{2}\eta_{ce}\right)\mathbf{e}^{c}\wedge\mathbf{e}^{d}\\
 & = & \mathbf{C}_{\;b}^{a}-2\Delta_{db}^{ae}\left(\mathcal{R}_{ec}-\phi_{e;c}+\phi_{e}\phi_{c}-\frac{1}{2}\phi^{2}\eta_{ce}\right)\mathbf{e}^{c}\wedge\mathbf{e}^{d}
\end{eqnarray*}
Equality of the traceless and trace parts shows immediately that both
\begin{eqnarray*}
\tilde{\mathbf{C}}_{\;b}^{a} & = & \mathbf{C}_{\;b}^{a}\\
\tilde{\boldsymbol{\mathcal{R}}}_{e} & = & e^{-\phi}\left(\mathcal{R}_{ec}-\phi_{e;c}+\phi_{e}\phi_{c}-\frac{1}{2}\phi^{2}\eta_{ce}\right)\mathbf{\tilde{e}}^{c}
\end{eqnarray*}
so the Weyl curvature 2-form is conformally invariant. The components
of each part transform as
\begin{eqnarray}
\tilde{C}_{\;bcd}^{a} & = & e^{-2\phi}C_{\;bcd}^{a}\label{Weyl curvature under conformal transformation}\\
\tilde{\mathcal{R}}_{ab} & = & e^{-2\phi}\left(\mathcal{R}_{ab}-\phi_{a;b}+\phi_{a}\phi_{b}-\frac{1}{2}\phi^{2}\eta_{ab}\right)\label{Schouten tensor under conformal transformation}\\
\tilde{\mathcal{R}} & = & e^{-2\phi}\left(\mathcal{R}-\phi_{\;;a}^{a}-\frac{1}{2}\left(n-2\right)\phi^{a}\phi_{a}\right)
\end{eqnarray}
where the factor of $e^{-2\phi}$ comes from replacing $\tilde{\mathbf{e}}^{a}=e^{\phi}\mathbf{e}^{a}$.
This proves that the Weyl curvature tensor is covariant with weight
$-2$ under a conformal transformation of the metric, and yields the
expression for the change in the Schouten (and therefore, Ricci) tensor
under conformal transformation.

\subsection{Conditions for conformal Ricci flatness}

Next, we find the condition required for the metric of a Riemannian
geometry to be conformally related to the metric of a Ricci-flat spacetime.
This follows as a pair of integrability conditions for $\phi$ when
we set eq.(\ref{Schouten tensor under conformal transformation})
equal to zero.

First, we rewrite eq.(\ref{Schouten tensor under conformal transformation})
as a $1$-form equation,
\[
\tilde{\boldsymbol{\mathcal{R}}}_{c}=e^{-\phi}\left(\boldsymbol{\mathcal{R}}_{c}-\mathbf{D}\phi_{c}+\phi_{c}\mathbf{d}\phi-\frac{1}{2}\eta^{ab}\phi_{a}\phi_{b}\eta_{ce}\mathbf{e}^{e}\right)
\]
where $\mathbf{D}\phi_{c}=\mathbf{d}\phi_{c}-\phi_{e}\boldsymbol{\omega}_{\;c}^{e}$.
Using the vector field $\phi_{c}\equiv e_{c}^{\quad\mu}\partial_{\mu}\phi$,
we define the corresponding $1$-form $\boldsymbol{\chi}\equiv\phi_{c}\mathbf{e}^{c}=\mathbf{d}\phi$
and ask for the conditions under which $\tilde{\boldsymbol{\mathcal{R}}}_{c}=0$
has a solution for $\boldsymbol{\phi}$. This may be written as a
pair of equations,
\begin{eqnarray}
\mathbf{d}\phi_{c} & = & \boldsymbol{\mathcal{R}}_{c}+\phi_{e}\boldsymbol{\omega}_{\;c}^{e}+\phi_{c}\boldsymbol{\chi}-\frac{1}{2}\eta^{ab}\phi_{a}\phi_{b}\eta_{ce}\mathbf{e}^{e}\label{Condition on phi for conformal Ricci flatness}\\
\mathbf{d}\phi & = & \boldsymbol{\chi}\label{Definition of phi}
\end{eqnarray}
The integrability conditions follow from the Poincaré lemma, $\mathbf{d}^{2}\equiv0$,
\begin{eqnarray}
0 & \equiv & \mathbf{d}^{2}\phi_{c}\nonumber \\
 & = & \mathbf{d}\boldsymbol{\mathcal{R}}_{c}+\mathbf{d}\phi_{e}\wedge\boldsymbol{\omega}_{\;c}^{e}+\phi_{e}\mathbf{d}\boldsymbol{\omega}_{\;c}^{e}+\mathbf{d}\phi_{c}\wedge\boldsymbol{\chi}-\eta^{ab}\phi_{a}\mathbf{d}\phi_{b}\wedge\eta_{ce}\mathbf{e}^{e}-\frac{1}{2}\phi^{2}\eta_{ce}\mathbf{d}\mathbf{e}^{e}\label{Integrability of vanishing Schouten}\\
0 & \equiv & \mathbf{d}^{2}\phi\nonumber \\
 & = & \mathbf{d}\boldsymbol{\chi}\label{Integrability of phi}
\end{eqnarray}
The second condition, eq.(\ref{Integrability of phi}), is identically
satisfied by the definition of $\boldsymbol{\chi}$. Substituting
the original equation for $\mathbf{d}\phi_{c}$, eq.(\ref{Condition on phi for conformal Ricci flatness}),
into the first integrability conditon, eq.(\ref{Integrability of vanishing Schouten}),
\begin{eqnarray*}
0 & = & \mathbf{d}\boldsymbol{\mathcal{R}}_{c}+\left(\boldsymbol{\mathcal{R}}_{e}+\phi_{d}\boldsymbol{\omega}_{\;e}^{d}+\phi_{e}\boldsymbol{\chi}-\frac{1}{2}\phi^{2}\eta_{ed}\mathbf{e}^{d}\right)\wedge\boldsymbol{\omega}_{\;c}^{e}+\phi_{e}\mathbf{d}\boldsymbol{\omega}_{\;c}^{e}\\
 &  & +\left(\boldsymbol{\mathcal{R}}_{c}+\phi_{e}\boldsymbol{\omega}_{\;c}^{e}+\phi_{c}\boldsymbol{\chi}-\frac{1}{2}\phi^{2}\eta_{ce}\mathbf{e}^{e}\right)\wedge\boldsymbol{\chi}\\
 &  & -\eta^{ab}\phi_{a}\left(\boldsymbol{\mathcal{R}}_{b}+\phi_{d}\boldsymbol{\omega}_{\;b}^{d}+\phi_{b}\boldsymbol{\chi}-\frac{1}{2}\phi^{2}\eta_{bd}\mathbf{e}^{d}\right)\wedge\eta_{ce}\mathbf{e}^{e}-\frac{1}{2}\phi^{2}\eta_{ce}\mathbf{d}\mathbf{e}^{e}\\
 & = & \left(\mathbf{d}\boldsymbol{\mathcal{R}}_{c}+\boldsymbol{\mathcal{R}}_{e}\wedge\boldsymbol{\omega}_{\;c}^{e}\right)+\left(\boldsymbol{\mathcal{R}}_{c}\wedge\boldsymbol{\chi}-\eta^{ab}\phi_{a}\boldsymbol{\mathcal{R}}_{b}\wedge\eta_{ce}\mathbf{e}^{e}\right)+\left(\phi_{e}\mathbf{d}\boldsymbol{\omega}_{\;c}^{e}+\phi_{d}\boldsymbol{\omega}_{\;e}^{d}\wedge\boldsymbol{\omega}_{\;c}^{e}\right)\\
 &  & -\frac{1}{2}\phi^{2}\left(\eta_{ce}\mathbf{d}\mathbf{e}^{e}+\eta_{ed}\mathbf{e}^{d}\wedge\boldsymbol{\omega}_{\;c}^{e}\right)-\frac{1}{2}\phi^{2}\eta_{ce}\mathbf{e}^{e}\wedge\boldsymbol{\chi}-\eta^{ab}\phi_{a}\phi_{d}\boldsymbol{\omega}_{\;b}^{d}\wedge\eta_{ce}\mathbf{e}^{e}\\
 &  & -\phi^{2}\boldsymbol{\chi}\wedge\eta_{ce}\mathbf{e}^{e}+\frac{1}{2}\phi^{2}\phi_{a}\mathbf{e}^{a}\wedge\eta_{ce}\mathbf{e}^{e}\\
 & = & \mathbf{D}\boldsymbol{\mathcal{R}}_{c}+\phi_{a}\left(\delta_{c}^{b}\delta_{e}^{a}-\eta^{ab}\eta_{ce}\right)\boldsymbol{\mathcal{R}}_{b}\wedge\mathbf{e}^{e}+\phi_{e}\mathbf{R}_{\;c}^{e}-\frac{1}{2}\phi^{2}\eta_{ec}\left(\mathbf{D}\mathbf{e}^{e}-\mathbf{e}^{d}\wedge\boldsymbol{\omega}_{\;d}^{e}\right)\\
 &  & +\phi^{2}\left(\frac{1}{2}\eta_{ce}\boldsymbol{\chi}\wedge\mathbf{e}^{e}-\eta_{ce}\boldsymbol{\chi}\wedge\mathbf{e}^{e}+\frac{1}{2}\eta_{ce}\boldsymbol{\chi}\wedge\mathbf{e}^{e}\right)-\left(\phi_{a}\phi_{d}\boldsymbol{\omega}^{da}\right)\wedge\eta_{ce}\mathbf{e}^{e}\\
 & = & \mathbf{D}\boldsymbol{\mathcal{R}}_{c}+\phi_{a}\left(\mathbf{R}_{\;c}^{a}+2\Delta_{ec}^{ab}\boldsymbol{\mathcal{R}}_{b}\wedge\mathbf{e}^{e}\right)
\end{eqnarray*}
which we see from eq.(\ref{Decomposition of the Riemann curvature})
becomes
\begin{equation}
0=\mathbf{D}\boldsymbol{\mathcal{R}}_{c}+\phi_{a}\mathbf{C}_{\;c}^{a}\label{Condition for conformal Ricci flatness}
\end{equation}
Though this well-known condition still depends on the gradient of
the conformal factor, $\phi_{a}$, Szekeres has shown using spinor
techniques that it can be broken down into two integrability conditions
depending only on the curvature \cite{Szekeres 1963}.

\subsection{Conditions for conformal Einstein equation with matter}

We may apply the same approach to the Einstein equation with conformal
matter. Let the matter be of definite conformal weight, $\tilde{\Psi}\rightarrow e^{k\phi}\Psi$
for a generic field $\Psi$. Then the covariant form of the stress-energy
tensor will be of conformal weight $-2$,
\[
\tilde{T}_{ab}=e^{-2\phi}T_{ab}
\]
to have the correct weight for the Einstein equation. The Einstein
tensor, of course, is not of definite conformal weight, but it acquires
an overall factor of $e^{-2\phi}$. We assume that $T_{ab}$ \emph{is}
of definite weight. 

Then, writing the Einstein equation, $R_{ab}-\frac{1}{2}\eta_{ab}R=T_{ab}$
in terms of the Schouten tensor using eqs.(\ref{Ricci in terms of Schouten})
and (\ref{Ricci scalar in terms of Schouten}), gives
\[
\mathcal{R}_{ab}-\eta_{ab}\mathcal{R}=\frac{1}{n-2}T_{ab}
\]
Now define, for arbitrary curvatures, not necessarily solutions,
\[
E_{ab}\equiv\mathcal{R}_{ab}-\eta_{ab}\mathcal{R}-\frac{1}{n-2}T_{ab}
\]
We would like to know when there exists a conformal transformation,
$E_{ab}\rightarrow\tilde{E}_{ab}$ such that $\tilde{E}_{ab}=0$.
The calculation is simpler if we notice that $E_{ab}=0$ if and only
if
\[
E_{ab}-\frac{1}{n-1}E\eta_{ab}=\mathcal{R}_{ab}-\frac{1}{n-2}\left(T_{ab}-\frac{1}{n-1}T\eta_{ab}\right)=0
\]
Defining
\begin{equation}
\mathbf{\mathcal{T}}_{ab}\equiv\frac{1}{n-2}\left(T_{ab}-\frac{1}{n-1}T\eta_{ab}\right)\label{Modified energy momentum tensor}
\end{equation}
we ask for a conformal gauge in which $\tilde{E}_{ab}-\frac{1}{n-1}\tilde{E}\eta_{ab}=\mathcal{R}_{ab}-\mathcal{T}_{ab}=0$.

We establish clearly that this is equivalent to the Einstein equation.
The essential question is the condition for a conformal transformation
such that $\tilde{E}_{ab}=0$. Substituting the conformally transformed
fields to find $\tilde{E}_{ab}$,
\begin{eqnarray*}
\tilde{E}_{ab} & = & \mathcal{R}_{ab}-\eta_{ab}\mathcal{R}-\frac{1}{n-2}T_{ab}\\
 & = & e^{-2\phi}\left(\mathcal{R}_{ab}-\phi_{a;b}+\phi_{a}\phi_{b}-\frac{1}{2}\phi^{2}\eta_{ab}\right)-\eta_{ab}e^{-2\phi}\left(\mathcal{R}-\phi_{\:;c}^{c}-\frac{1}{2}\left(n-2\right)\phi^{c}\phi_{c}\right)-\frac{1}{n-2}e^{-2\phi}T_{ab}
\end{eqnarray*}
so we examine integrability of
\begin{eqnarray}
0 & = & \left(\mathcal{R}_{ab}-\phi_{a;b}+\phi_{a}\phi_{b}-\frac{1}{2}\phi^{2}\eta_{ab}\right)-\eta_{ab}\left(\mathcal{R}-\phi_{\:;c}^{c}-\frac{1}{2}\left(n-2\right)\phi^{c}\phi_{c}\right)-\frac{1}{n-2}T_{ab}\label{Condition for phi to give vanishing E}
\end{eqnarray}
However, by solving the trace of this equation for $\phi_{\:;c}^{c}$,
\begin{eqnarray}
0 & = & \mathcal{R}-\phi_{\;;a}^{a}+\phi^{a}\phi_{a}-\frac{1}{2}n\phi^{c}\phi_{c}-\left(n\mathcal{R}-n\phi_{\:;c}^{c}-\frac{1}{2}n\left(n-2\right)\phi^{c}\phi_{c}\right)-\frac{1}{n-2}T\nonumber \\
 & = & -\left(n-1\right)\mathcal{R}+\left(n-1\right)\phi_{\:;c}^{c}+\frac{1}{2}\left(n-1\right)\left(n-2\right)\phi^{c}\phi_{c}-\frac{1}{n-2}T\nonumber \\
\phi_{\:;c}^{c} & = & \mathcal{R}-\frac{1}{2}\left(n-2\right)\phi^{c}\phi_{c}+\frac{1}{\left(n-1\right)\left(n-2\right)}T\label{Trace of condition for vanishing E}
\end{eqnarray}
and substituting eq.(\ref{Trace of condition for vanishing E}) back
into eq.(\ref{Condition for phi to give vanishing E}) we find the
simpler form,
\begin{eqnarray}
0 & = & \mathcal{R}_{ab}-\phi_{a;b}+\phi_{a}\phi_{b}-\frac{1}{2}\phi^{2}\eta_{ab}-\mathcal{T}_{ab}\label{Simplified condition for vanishing E}
\end{eqnarray}
and this is just $\tilde{E}_{ab}-\frac{1}{n-1}\tilde{E}\eta_{ab}=0$.
Conversely, the trace of eq.(\ref{Simplified condition for vanishing E})
reproduces the trace condition, eq.(\ref{Trace of condition for vanishing E}).
Therefore, conformal vanishing of $\tilde{E}_{ab}$ is equivalent
to conformal vanishing of $\tilde{E}_{ab}-\frac{1}{n-1}\tilde{E}\eta_{ab}$. 

Returning to the find the condition, we set $\boldsymbol{\mathcal{T}}_{a}\equiv\mathcal{T}_{ab}\mathbf{e}^{b}$
and $\boldsymbol{\chi}=\mathbf{d}\phi$, then write $\tilde{E}_{ab}-\frac{1}{n-1}\tilde{E}\eta_{ab}=0$
as a $1$-form equation,
\begin{eqnarray*}
0 & = & \left(\boldsymbol{\mathcal{R}}_{a}-\mathbf{d}\phi_{a}+\phi_{b}\boldsymbol{\omega}_{\:a}^{b}+\phi_{a}\boldsymbol{\chi}-\frac{1}{2}\phi^{2}\eta_{ab}\mathbf{e}^{b}\right)-\boldsymbol{\mathcal{T}}_{a}
\end{eqnarray*}
We therefore require
\begin{eqnarray*}
\mathbf{d}\phi_{a} & = & \boldsymbol{\mathcal{R}}_{a}+\phi_{b}\boldsymbol{\omega}_{\:a}^{b}+\phi_{a}\boldsymbol{\chi}-\frac{1}{2}\phi^{2}\eta_{ab}\mathbf{e}^{b}-\boldsymbol{\mathcal{T}}_{a}\\
\mathbf{d}\boldsymbol{\chi} & = & \mathbf{d}^{2}\phi\;\;\equiv\;\;0
\end{eqnarray*}
The trace relation of eq.(\ref{Trace of condition for vanishing E})
also holds. The integrability condition is:
\begin{eqnarray*}
0 & \equiv & \mathbf{d}^{2}\phi_{a}\\
 & = & \mathbf{d}\left(\boldsymbol{\mathcal{R}}_{a}-\boldsymbol{\mathcal{T}}_{a}+\phi_{b}\boldsymbol{\omega}_{\:a}^{b}+\phi_{a}\boldsymbol{\chi}-\frac{1}{2}\phi^{2}\eta_{ab}\mathbf{e}^{b}\right)\\
 & = & \mathbf{d}\left(\boldsymbol{\mathcal{R}}_{a}-\boldsymbol{\mathcal{T}}_{a}\right)+\phi_{b}\mathbf{d}\boldsymbol{\omega}_{\:a}^{b}-\frac{1}{2}\phi^{2}\eta_{ab}\mathbf{d}\mathbf{e}^{b}\\
 &  & +\left(\boldsymbol{\mathcal{R}}_{b}-\boldsymbol{\mathcal{T}}_{b}+\phi_{c}\boldsymbol{\omega}_{\:b}^{c}+\phi_{b}\boldsymbol{\chi}-\frac{1}{2}\phi^{2}\eta_{bc}\mathbf{e}^{c}\right)\wedge\boldsymbol{\omega}_{\:a}^{b}\\
 &  & +\left(\boldsymbol{\mathcal{R}}_{a}-\boldsymbol{\mathcal{T}}_{a}+\phi_{b}\boldsymbol{\omega}_{\:a}^{b}+\phi_{a}\boldsymbol{\chi}-\frac{1}{2}\phi^{2}\eta_{ab}\mathbf{e}^{b}\right)\wedge\boldsymbol{\chi}\\
 &  & -\phi^{c}\left(\boldsymbol{\mathcal{R}}_{c}-\boldsymbol{\mathcal{T}}_{c}+\phi_{b}\boldsymbol{\omega}_{\:c}^{b}+\phi_{c}\boldsymbol{\chi}-\frac{1}{2}\phi^{2}\eta_{cb}\mathbf{e}^{b}\right)\wedge\eta_{ad}\mathbf{e}^{d}
\end{eqnarray*}
Distributing and collecting like terms,
\begin{eqnarray*}
0 & = & \mathbf{d}\left(\boldsymbol{\mathcal{R}}_{a}-\boldsymbol{\mathcal{T}}_{a}\right)+\left(\boldsymbol{\mathcal{R}}_{b}-\boldsymbol{\mathcal{T}}_{b}\right)\wedge\boldsymbol{\omega}_{\:a}^{b}+\left(\boldsymbol{\mathcal{R}}_{a}-\boldsymbol{\mathcal{T}}_{a}\right)\wedge\mathbf{d}\phi-\phi^{c}\left(\boldsymbol{\mathcal{R}}_{c}-\boldsymbol{\mathcal{T}}_{c}\right)\wedge\eta_{ad}\mathbf{e}^{d}\\
 &  & +\phi_{b}\left(\mathbf{d}\boldsymbol{\omega}_{\:a}^{b}-\boldsymbol{\omega}_{\:a}^{c}\wedge\boldsymbol{\omega}_{\:c}^{b}\right)-\frac{1}{2}\phi^{2}\eta_{ab}\left(\mathbf{d}\mathbf{e}^{b}-\mathbf{e}^{c}\wedge\boldsymbol{\omega}_{\:c}^{b}\right)\\
 &  & +\left(\phi_{b}\boldsymbol{\chi}\wedge\boldsymbol{\omega}_{\:a}^{b}+\phi_{b}\boldsymbol{\omega}_{\:a}^{b}\wedge\boldsymbol{\chi}\right)+\left(\frac{1}{2}\phi^{c}\phi_{c}-\phi^{c}\phi_{c}+\frac{1}{2}\phi^{c}\phi_{c}\right)\mathbf{d}\phi\wedge\eta_{ad}\mathbf{e}^{d}\\
 & = & \mathbf{D}\left(\boldsymbol{\mathcal{R}}_{a}-\boldsymbol{\mathcal{T}}_{a}\right)+\phi_{b}\mathbf{R}_{\:a}^{b}+\phi_{b}\delta_{a}^{c}\delta_{d}^{b}\left(\boldsymbol{\mathcal{R}}_{c}-\boldsymbol{\mathcal{T}}_{c}\right)\wedge\mathbf{e}^{d}-\phi_{b}\eta^{bc}\eta_{ad}\left(\boldsymbol{\mathcal{R}}_{c}-\boldsymbol{\mathcal{T}}_{c}\right)\wedge\mathbf{e}^{d}\\
 & = & \mathbf{D}\left(\boldsymbol{\mathcal{R}}_{a}-\boldsymbol{\mathcal{T}}_{a}\right)+\phi_{b}\mathbf{R}_{\:a}^{b}+2\phi_{b}\Delta_{ad}^{cb}\left(\boldsymbol{\mathcal{R}}_{c}-\boldsymbol{\mathcal{T}}_{c}\right)\wedge\mathbf{e}^{d}\\
 & = & \mathbf{D}\boldsymbol{\mathcal{R}}_{a}+\phi_{b}\mathbf{C}_{\:a}^{b}-\mathbf{D}\boldsymbol{\mathcal{T}}_{a}-2\phi_{b}\Delta_{ad}^{cb}\boldsymbol{\mathcal{T}}_{c}\wedge\mathbf{e}^{d}
\end{eqnarray*}
leaving us with
\begin{eqnarray}
\mathbf{D}\boldsymbol{\mathcal{R}}_{a}+\phi_{b}\mathbf{C}_{\;a}^{b} & = & \mathbf{D}\boldsymbol{\mathcal{T}}_{a}+\phi_{b}2\Delta_{da}^{bc}\boldsymbol{\mathcal{T}}_{c}\mathbf{e}^{d}\label{Condition for conformal Einstein equation}
\end{eqnarray}
This is a new result. When eq.(\ref{Simplified condition for vanishing E})
is written using the Riemann tensor instead of the Weyl tensor,
\begin{eqnarray*}
\mathbf{D}\left(\boldsymbol{\mathcal{R}}_{a}-\boldsymbol{\mathcal{T}}_{a}\right)+\phi_{b}\mathbf{R}_{\:a}^{b}+2\phi_{b}\Delta_{ad}^{cb}\left(\boldsymbol{\mathcal{R}}_{c}-\boldsymbol{\mathcal{T}}_{c}\right)\wedge\mathbf{e}^{d} & = & 0
\end{eqnarray*}
we recognize the same condition as that for Ricci flatness, but with
the Schouten tensor replaced by $\boldsymbol{\mathcal{R}}_{a}-\boldsymbol{\mathcal{T}}_{a}$.

\section{Weyl geometry}

A simple extension of the Poincaré symmetry underlying Riemannian
geometry leads to the Cartan structure equations for the Weyl group:
\begin{eqnarray}
\boldsymbol{\mathfrak{R}}_{\;b}^{a} & = & \mathbf{d}\boldsymbol{\omega}_{\;b}^{a}-\boldsymbol{\omega}_{\;b}^{c}\wedge\boldsymbol{\omega}_{\;c}^{a}\label{Structure equation for Weyl spin connection}\\
\mathbf{T}^{a} & = & \mathbf{d}\mathbf{e}^{a}-\mathbf{e}^{b}\wedge\boldsymbol{\omega}_{\;b}^{a}-\boldsymbol{\omega}\wedge\mathbf{e}^{a}\label{Structure equation for Weyl solder form}\\
\boldsymbol{\Omega} & = & \mathbf{d}\boldsymbol{\omega}\label{Structure equation for Weyl vector}
\end{eqnarray}
where the most general case includes both the torsion, $\mathbf{T}^{a}=\frac{1}{2}T_{\;bc}^{a}\mathbf{e}^{b}\wedge\mathbf{e}^{c}$,
and the dilatational curvature, $\boldsymbol{\Omega}=\frac{1}{2}\Omega_{ab}\mathbf{e}^{b}\wedge\mathbf{e}^{c}$.
In our treatment of a Dirac-like theory, we will not assume vanishing
torsion.

A conformal transformation of the metric, eq.(\ref{Conformal transformation}),
now transforms both the solder form and the Weyl vector, according
to
\begin{eqnarray*}
\tilde{\mathbf{e}}^{a} & = & e^{\phi}\mathbf{e}^{a}\\
\tilde{\boldsymbol{\omega}} & = & \boldsymbol{\omega}+\mathbf{d}\phi
\end{eqnarray*}
The final structure equation, eq.(\ref{Structure equation for Weyl vector})
then remains unchanged, since
\[
\mathbf{d}\tilde{\boldsymbol{\omega}}=\mathbf{d}\boldsymbol{\omega}
\]
The basis equation transforms as
\begin{eqnarray*}
\tilde{\mathbf{T}}^{a} & = & \mathbf{d}\tilde{\mathbf{e}}^{a}-\tilde{\mathbf{e}}^{b}\wedge\tilde{\boldsymbol{\omega}}_{\;b}^{a}-\tilde{\boldsymbol{\omega}}\wedge\tilde{\mathbf{e}}^{a}\\
 & = & \left(e^{\phi}\mathbf{d}\phi\wedge\mathbf{e}^{a}+e^{\phi}\mathbf{d}\mathbf{e}^{a}\right)-e^{\phi}\mathbf{e}^{b}\wedge\tilde{\boldsymbol{\omega}}_{\;b}^{a}-\left(\boldsymbol{\omega}+\mathbf{d}\phi\right)\wedge e^{\phi}\mathbf{e}^{a}\\
 & = & e^{\phi}\left(\mathbf{d}\phi\wedge\mathbf{e}^{a}+\left(\mathbf{T}^{a}+\mathbf{e}^{c}\wedge\boldsymbol{\omega}_{\;c}^{a}+\boldsymbol{\omega}\wedge\mathbf{e}^{a}\right)-\mathbf{e}^{b}\wedge\tilde{\boldsymbol{\omega}}_{\;b}^{a}-\boldsymbol{\omega}\wedge\mathbf{e}^{a}-\mathbf{d}\phi\wedge\mathbf{e}^{a}\right)\\
 & = & e^{\phi}\mathbf{T}^{a}+e^{\phi}\mathbf{e}^{b}\wedge\left(\boldsymbol{\omega}_{\;b}^{a}-\tilde{\boldsymbol{\omega}}_{\;b}^{a}\right)
\end{eqnarray*}
We conclude that it is sufficient to take the spin connection to be
conformally invariant, and the torsion a weight-$1$ conformal tensor:
\begin{eqnarray*}
\tilde{\boldsymbol{\omega}}_{\;b}^{a} & = & \boldsymbol{\omega}_{\;b}^{a}\\
\tilde{\mathbf{T}}^{a} & = & e^{\phi}\mathbf{T}^{a}
\end{eqnarray*}
These inferences are correct, as may be shown directly from the gauge
transformation properties of the Cartan connection. Since the spin
connection is invariant, the Lorentz curvature $2$-form is also invariant,
$\tilde{\boldsymbol{\mathfrak{R}}}_{\;b}^{a}=\boldsymbol{\mathfrak{R}}_{\;b}^{a}$.

We again use the Poincaré lemma, $\mathbf{d}^{2}\equiv0$ to find
the integrability conditions:
\begin{eqnarray}
\mathbf{D}\boldsymbol{\mathfrak{R}}_{\;b}^{a} & = & 0\label{Curvature Bianchi}\\
\mathbf{D}\mathbf{T}^{a} & = & \mathbf{e}^{b}\wedge\boldsymbol{\mathfrak{R}}_{\;b}^{a}-\boldsymbol{\Omega}\wedge\mathbf{e}^{a}\label{Torsion Bianchi}\\
\mathbf{D}\boldsymbol{\Omega} & = & 0\label{Dilatation Bianchi}
\end{eqnarray}
where the covariant derivatives are given by
\begin{eqnarray*}
\mathbf{D}\boldsymbol{\mathfrak{R}}_{\;b}^{a} & \equiv & \mathbf{d}\boldsymbol{\mathfrak{R}}_{\;b}^{a}+\boldsymbol{\mathfrak{R}}_{\;b}^{c}\wedge\boldsymbol{\omega}_{\;c}^{a}-\boldsymbol{\mathfrak{R}}_{\;c}^{a}\wedge\boldsymbol{\omega}_{\;b}^{c}\\
\mathbf{D}\mathbf{T}^{a} & \equiv & \mathbf{d}\mathbf{T}^{a}+\mathbf{T}^{b}\wedge\boldsymbol{\omega}_{\;b}^{a}-\boldsymbol{\omega}\wedge\mathbf{T}^{a}\\
\mathbf{D}\boldsymbol{\Omega} & \equiv & \mathbf{d}\boldsymbol{\Omega}
\end{eqnarray*}

When the torsion vanishes, we have a pair of algebraic identities
since the Weyl-Ricci tensor may have an antisymmetric part. From 
\begin{eqnarray}
\mathbf{e}^{b}\wedge\boldsymbol{\mathfrak{R}}_{\;b}^{a} & = & \boldsymbol{\Omega}\wedge\mathbf{e}^{a}\nonumber \\
\mathfrak{R}_{\;\left[bcd\right]}^{a} & = & \delta_{[b}^{a}\Omega_{cd]}\label{Vanishing torsion Bianchi}
\end{eqnarray}
we find the symmetric and antisymmetric parts,
\begin{eqnarray*}
\mathfrak{R}_{\;bcd}^{a}+\mathfrak{R}_{\;cdb}^{a}+\mathfrak{R}_{\;dbc}^{a} & = & \delta_{b}^{a}\Omega_{cd}+\delta_{c}^{a}\Omega_{db}+\delta_{d}^{a}\Omega_{bc}\\
\mathfrak{R}_{bd}-\mathfrak{R}_{db} & = & -\left(n-2\right)\Omega_{bd}
\end{eqnarray*}
While the Lorentz curvature $2$-form is conformally invariant, the
components $\mathfrak{R}_{\;bcd}^{a},\mathfrak{R}_{ab}$ and $\Omega_{ab}$
all have conformal weight $-2$.

\subsection{The connection with the Weyl vector and torsion}

As with Riemannian geometry, the structure equation for the solder
form, eq.(\ref{Structure equation for Weyl solder form}). allows
us to solve for the connection.

\subsubsection{Weyl connection with torsion}

Look at the solder form equation,
\[
\mathbf{d}\mathbf{e}^{a}=\mathbf{e}^{b}\wedge\boldsymbol{\omega}_{\;b}^{a}+\boldsymbol{\omega}\wedge\mathbf{e}^{a}+\mathbf{T}^{a}
\]
Notice that when $\mathbf{T}^{a}=0$ this has exactly the same form
as the conformally transformed solder form structure equation of a
Riemannian geometry, eq.(\ref{Structure equation for conformally transformed spacetime}),
with $\mathbf{d}\phi$ replaced by $-\boldsymbol{\omega}$. Thus,
the solution for the Weyl spin connection is completely analogous
to the effect of a dilatation on the connection of a Riemannian geometry,
with the Weyl vector $W_{c}$ replacing the negative of the gradient
of the scale change, $-\phi_{c}$ in eq.(\ref{Transformation of the spin connection}).
Taking advantage of this observation, let $\boldsymbol{\omega}_{\;b}^{a}=\boldsymbol{\alpha}_{\;b}^{a}+\boldsymbol{\beta}_{\;b}^{a}+\boldsymbol{\gamma}_{\;b}^{a}$
where $\boldsymbol{\alpha}_{\;b}^{a}$ is the compatible connection
and $\boldsymbol{\beta}_{\;b}^{a}$ is the required Weyl vector piece,
\begin{eqnarray*}
\mathbf{d}\mathbf{e}^{a} & = & \mathbf{e}^{b}\wedge\boldsymbol{\alpha}_{\;b}^{a}\\
\boldsymbol{\beta}_{\;b}^{a} & = & -2\Delta_{db}^{ac}W_{c}\mathbf{e}^{d}
\end{eqnarray*}
and each term has the same antisymmetry of indices as the full spin
connection, i.e., $\boldsymbol{\omega}_{\;b}^{a}=-\eta^{ac}\eta_{bd}\boldsymbol{\omega}_{\;c}^{d}$.
Then
\begin{eqnarray*}
\mathbf{d}\mathbf{e}^{a} & = & \mathbf{e}^{b}\wedge\left(\boldsymbol{\alpha}_{\;b}^{a}+\boldsymbol{\beta}_{\;b}^{a}+\boldsymbol{\gamma}_{\;b}^{a}\right)+\boldsymbol{\omega}\wedge\mathbf{e}^{a}+\mathbf{T}^{a}\\
 & = & \mathbf{e}^{b}\wedge\boldsymbol{\alpha}_{\;b}^{a}+\mathbf{e}^{b}\wedge\boldsymbol{\beta}_{\;b}^{a}+\mathbf{e}^{b}\wedge\boldsymbol{\gamma}_{\;b}^{a}+\boldsymbol{\omega}\wedge\mathbf{e}^{a}+\mathbf{T}^{a}\\
0 & = & \left(-2\Delta_{db}^{ac}W_{c}\mathbf{e}^{b}\wedge\mathbf{e}^{d}+\boldsymbol{\omega}\wedge\mathbf{e}^{a}\right)+\left(\mathbf{e}^{b}\wedge\boldsymbol{\gamma}_{\;b}^{a}+\mathbf{T}^{a}\right)\\
 & = & \mathbf{e}^{b}\wedge\boldsymbol{\gamma}_{\;b}^{a}+\mathbf{T}^{a}\\
 & = & \left(\gamma_{\;\;bc}^{a}+\frac{1}{2}T_{\;\;bc}^{a}\right)\mathbf{e}^{b}\wedge\mathbf{e}^{c}
\end{eqnarray*}
The final equation must involve antisymmetric $\eta_{ae}\gamma_{\;\;bc}^{e}=\gamma_{abc}=-\gamma_{bac}$.
Lowering indices in the remaining condition and cycling,
\begin{eqnarray*}
0 & = & \gamma_{abc}-\gamma_{acb}+T_{abc}\\
0 & = & \gamma_{bca}-\gamma_{bac}+T_{bca}\\
0 & = & \gamma_{cab}-\gamma_{cba}+T_{cab}
\end{eqnarray*}
we combine with the usual sum-sum-difference and solve.
\begin{eqnarray*}
0 & = & \gamma_{abc}-\gamma_{acb}+\gamma_{bca}-\gamma_{bac}-\gamma_{cab}+\gamma_{cba}+\left(T_{abc}-T_{cab}+T_{bca}\right)\\
 & = & \left(\gamma_{abc}-\gamma_{bac}\right)-\left(\gamma_{acb}+\gamma_{cab}\right)+\left(\gamma_{bca}+\gamma_{cba}\right)+\left(T_{abc}-T_{cab}+T_{bca}\right)\\
 & = & 2\gamma_{abc}+\left(T_{abc}-T_{cab}+T_{bca}\right)\\
\gamma_{abc} & = & -\frac{1}{2}\left(T_{abc}-T_{cab}+T_{bca}\right)
\end{eqnarray*}
Therefore,
\[
\gamma_{\;\;bc}^{a}=-\frac{1}{2}\left(T_{\;\;bc}^{a}+T_{cb}^{\quad a}+T_{bc}^{\quad a}\right)
\]
and the spin connection is given by
\begin{eqnarray}
\boldsymbol{\omega}_{\;b}^{a} & = & \boldsymbol{\alpha}_{\;b}^{a}-2\Delta_{db}^{ac}W_{c}\mathbf{e}^{d}-C_{\;\;bc}^{a}\mathbf{e}^{c}\label{Weyl spin connection with torsion}
\end{eqnarray}
where we define the contorsion tensor to be
\begin{equation}
C_{\;\;bc}^{a}\equiv\frac{1}{2}\left(T_{\;\;bc}^{a}+T_{cb}^{\quad a}+T_{bc}^{\quad a}\right)\label{Contorsion tensor-1}
\end{equation}
Now check,
\begin{eqnarray*}
\mathbf{d}\mathbf{e}^{a} & = & \mathbf{e}^{b}\wedge\boldsymbol{\omega}_{\;b}^{a}+\boldsymbol{\omega}\wedge\mathbf{e}^{a}+\mathbf{T}^{a}\\
 & = & \mathbf{e}^{b}\wedge\left(\boldsymbol{\alpha}_{\;b}^{a}-2\Delta_{db}^{ac}W_{c}\mathbf{e}^{d}-C_{\;\;bc}^{a}\mathbf{e}^{c}\right)+\boldsymbol{\omega}\wedge\mathbf{e}^{a}+\mathbf{T}^{a}\\
 & = & \mathbf{e}^{b}\wedge\boldsymbol{\alpha}_{\;b}^{a}-\left(\delta_{d}^{a}\delta_{b}^{c}-\eta^{ac}\eta_{bd}\right)W_{c}\mathbf{e}^{b}\wedge\mathbf{e}^{d}-C_{\;\;bc}^{a}\mathbf{e}^{b}\wedge\mathbf{e}^{c}+\boldsymbol{\omega}\wedge\mathbf{e}^{a}+\mathbf{T}^{a}\\
 & = & \mathbf{e}^{b}\wedge\boldsymbol{\alpha}_{\;b}^{a}-W_{b}\mathbf{e}^{b}\wedge\mathbf{e}^{a}+\boldsymbol{\omega}\wedge\mathbf{e}^{a}-\frac{1}{2}\left(T_{\;\;bc}^{a}+T_{cb}^{\quad a}+T_{bc}^{\quad a}\right)\mathbf{e}^{b}\wedge\mathbf{e}^{c}+\frac{1}{2}T_{\;\;bc}^{a}\mathbf{e}^{b}\wedge\mathbf{e}^{c}\\
 & = & \mathbf{e}^{b}\wedge\boldsymbol{\alpha}_{\;b}^{a}-\frac{1}{2}T_{\;\;bc}^{a}\mathbf{e}^{b}\wedge\mathbf{e}^{c}+\frac{1}{2}T_{\;\;bc}^{a}\mathbf{e}^{b}\wedge\mathbf{e}^{c}\\
 & = & \mathbf{e}^{b}\wedge\boldsymbol{\alpha}_{\;b}^{a}
\end{eqnarray*}

\subsubsection{The covariant derivative of Weyl geometry in a coordinate basis}

When a tensor transforms linearly and homogeneously with a power $\lambda$
of the conformal transformation that applies to the solder form $\tilde{\mathbf{e}}^{a}=e^{\phi}\mathbf{e}^{a}$,
\[
\tilde{T}^{A}=e^{\lambda\phi}T^{A}
\]
it is a \emph{conformal tensor of weight} $\lambda$. When differentiating
a conformal tensor of weight $\lambda$ the covariant derivative in
Weyl geometry is not just the partial derivative, but includes the
weight of the field times the field, times the Weyl vector. For example,
for a scalar field we have
\[
D_{\mu}\varphi=\partial_{\mu}\varphi-\lambda W_{\mu}\varphi
\]
This means that metric compatibility gives a different expression
for the connection.
\begin{eqnarray*}
0 & = & D_{\mu}g_{\alpha\beta}\\
 & = & \partial_{\mu}g_{\alpha\beta}-g_{\nu\beta}\tilde{\Gamma}_{\;\alpha\mu}^{\nu}-g_{\alpha\nu}\hat{\Gamma}_{\;\beta\mu}^{\nu}-2W_{\mu}g_{\alpha\beta}\\
 & = & \partial_{\mu}g_{\alpha\beta}-\hat{\Gamma}_{\beta\alpha\mu}-\hat{\Gamma}_{\alpha\beta\mu}-2W_{\mu}g_{\alpha\beta}
\end{eqnarray*}

The derivative of a contravariant vector of weight $\lambda$ is given
by
\begin{eqnarray}
D_{\mu}v^{\alpha} & = & \partial_{\mu}v^{\alpha}+v^{\beta}\hat{\Gamma}_{\;\beta\mu}^{\alpha}-\lambda v^{\alpha}W_{\mu}\label{Covariant derivative of weight lambda vector}
\end{eqnarray}
and, checking the tranformed derivative,
\begin{eqnarray*}
\tilde{D}_{\mu}\tilde{v}^{\alpha} & = & \tilde{D}_{\mu}\left(e^{\lambda\phi}v^{\alpha}\right)\\
 & = & \partial_{\mu}\left(e^{\lambda\phi}v^{\alpha}\right)+e^{\lambda\phi}v^{\beta}\tilde{\Gamma}_{\;\beta\mu}^{\alpha}-\lambda e^{\lambda\phi}v^{\alpha}\left(W_{\mu}+\partial_{\mu}\phi\right)\\
 & = & e^{\lambda\phi}\left(\partial_{\mu}v^{\alpha}+v^{\beta}\tilde{\Gamma}_{\;\beta\mu}^{\alpha}-\lambda v^{\alpha}W_{\mu}\right)\\
 & = & e^{\lambda\phi}D_{\mu}v^{\alpha}
\end{eqnarray*}
and is therefore properly covariant. Notice that the Weyl connection
is invariant under a conformal transformation, $\tilde{\Gamma}_{\;\beta\mu}^{\alpha}=\Gamma_{\;\beta\mu}^{\alpha}$.

\subsubsection{Weyl connection with torsion in a coordinate basis}

The corresponding expression in a coordinate basis starts with the
definition of torsion as the antisymmetric part of the connection,
\begin{eqnarray}
T_{\alpha\mu\beta} & \equiv & \hat{\Gamma}_{\alpha\mu\beta}-\hat{\Gamma}_{\alpha\beta\mu}\label{Torsion}
\end{eqnarray}
Then, starting from metric compatibility,
\begin{eqnarray*}
0 & = & D_{\mu}g_{\alpha\beta}\\
 & = & \partial_{\mu}g_{\alpha\beta}-g_{\nu\beta}\tilde{\Gamma}_{\;\alpha\mu}^{\nu}-g_{\alpha\nu}\hat{\Gamma}_{\;\beta\mu}^{\nu}-2W_{\mu}g_{\alpha\beta}\\
 & = & \partial_{\mu}g_{\alpha\beta}-\hat{\Gamma}_{\beta\alpha\mu}-\hat{\Gamma}_{\alpha\beta\mu}-2W_{\mu}g_{\alpha\beta}
\end{eqnarray*}
we cycle the expression in the usual way
\begin{eqnarray*}
\hat{\Gamma}_{\beta\alpha\mu}+\hat{\Gamma}_{\alpha\beta\mu} & = & \partial_{\mu}g_{\alpha\beta}-2W_{\mu}g_{\alpha\beta}\\
\hat{\Gamma}_{\alpha\mu\beta}+\hat{\Gamma}_{\mu\alpha\beta} & = & \partial_{\beta}g_{\mu\alpha}-2W_{\beta}g_{\mu\alpha}\\
\hat{\Gamma}_{\mu\beta\alpha}+\hat{\Gamma}_{\beta\mu\alpha} & = & \partial_{\alpha}g_{\beta\mu}-2W_{\alpha}g_{\beta\mu}
\end{eqnarray*}
Each of these three expressions is a conformal tensor since
\begin{eqnarray*}
\partial_{\mu}\tilde{g}_{\alpha\beta}-2\tilde{W}_{\mu}\tilde{g}_{\alpha\beta} & = & \partial_{\mu}\left(e^{2\phi}g_{\alpha\beta}\right)-2\left(W_{\mu}+\partial_{\mu}\phi\right)e^{2\phi}g_{\alpha\beta}\\
 & = & e^{2\phi}\left(\partial_{\mu}g_{\alpha\beta}-2W_{\mu}g_{\alpha\beta}\right)
\end{eqnarray*}
Adding the first two and subtracting the third we no longer assume
the connection is symmetric,
\begin{eqnarray*}
0 & = & \hat{\Gamma}_{\beta\alpha\mu}+\hat{\Gamma}_{\alpha\beta\mu}+\hat{\Gamma}_{\alpha\mu\beta}+\hat{\Gamma}_{\mu\alpha\beta}-\hat{\Gamma}_{\mu\beta\alpha}-\hat{\Gamma}_{\beta\mu\alpha}\\
 &  & -\partial_{\mu}g_{\alpha\beta}+2W_{\mu}g_{\alpha\beta}-\partial_{\beta}g_{\mu\alpha}+2W_{\beta}g_{\mu\alpha}+\partial_{\alpha}g_{\beta\mu}-2W_{\alpha}g_{\beta\mu}\\
 & = & 2\hat{\Gamma}_{\alpha\beta\mu}+\left(\hat{\Gamma}_{\alpha\mu\beta}-\hat{\Gamma}_{\alpha\beta\mu}\right)+\left(\hat{\Gamma}_{\beta\alpha\mu}-\hat{\Gamma}_{\beta\mu\alpha}\right)+\left(\hat{\Gamma}_{\mu\alpha\beta}-\hat{\Gamma}_{\mu\beta\alpha}\right)\\
 &  & -\left(\partial_{\mu}g_{\alpha\beta}+\partial_{\beta}g_{\mu\alpha}-\partial_{\alpha}g_{\beta\mu}\right)+2\left(W_{\mu}g_{\alpha\beta}+W_{\beta}g_{\mu\alpha}-W_{\alpha}g_{\beta\mu}\right)\\
 & = & 2\hat{\Gamma}_{\alpha\beta\mu}+\left(T_{\alpha\mu\beta}+T_{\beta\alpha\mu}+T_{\mu\alpha\beta}\right)\\
 &  & -\left(\partial_{\mu}g_{\alpha\beta}+\partial_{\beta}g_{\mu\alpha}-\partial_{\alpha}g_{\beta\mu}\right)+2\left(W_{\mu}g_{\alpha\beta}+W_{\beta}g_{\mu\alpha}-W_{\alpha}g_{\beta\mu}\right)
\end{eqnarray*}
we find
\begin{eqnarray*}
\hat{\Gamma}_{\alpha\beta\mu} & = & \Gamma_{\alpha\beta\mu}-\left(W_{\mu}g_{\alpha\beta}+W_{\beta}g_{\mu\alpha}-W_{\alpha}g_{\beta\mu}\right)-\frac{1}{2}\left(T_{\alpha\mu\beta}+T_{\beta\alpha\mu}+T_{\mu\alpha\beta}\right)
\end{eqnarray*}
Now, if we raise the first index,
\begin{align}
\hat{\Gamma}_{\;\;\beta\mu}^{\alpha} & =\Gamma_{\;\;\beta\mu}^{\alpha}-\left(\delta_{\beta}^{\alpha}W_{\mu}+\delta_{\mu}^{\alpha}W_{\beta}-W^{\alpha}g_{\beta\mu}\right)-\frac{1}{2}\left(T_{\;\;\mu\beta}^{\alpha}+T_{\beta\quad\mu}^{\;\;\;\alpha}+T_{\mu\quad\beta}^{\;\;\;\alpha}\right)\label{Weyl connection in coordinate basis with torsion}
\end{align}
we arrive at the coordinate form of the connection.

\subsection{The Weyl-Schouten tensor}

The invariance of the full curvature, $\tilde{\boldsymbol{\mathfrak{R}}}_{\;b}^{a}=\boldsymbol{\mathfrak{R}}_{\;b}^{a}$,
means that not only is the Weyl curvature of a Weyl geometry conformally
covariant, but so is the Weyl-Schouten tensor, $\mathscr{R}_{a}$.
By the Weyl-Schouten tensor, we mean the conformally covariant Schouten
tensor of a Weyl geometry. To compute it for a torsion-free Weyl geometry,
we must expand the curvature, eq.(\ref{Structure equation for Weyl spin connection})
using the Weyl connection,
\begin{equation}
\boldsymbol{\omega}_{\;b}^{a}=\boldsymbol{\alpha}_{\;b}^{a}-2\Delta_{db}^{ac}W_{c}\mathbf{e}^{d}\label{Weyl connection}
\end{equation}
with $\boldsymbol{\alpha}_{\;b}^{a}$ still the metric-compatible
spin connection. The difference is that now \emph{all} of $\boldsymbol{\mathfrak{R}}_{\;b}^{a}$
will be conformally covariant because the connection is scale invariant,
\begin{eqnarray*}
\mathbf{d}\tilde{\mathbf{e}}^{a} & = & \tilde{\mathbf{e}}^{b}\wedge\tilde{\boldsymbol{\omega}}_{\;b}^{a}+\tilde{\boldsymbol{\omega}}\wedge\tilde{\mathbf{e}}^{a}\\
\mathbf{d}\left(e^{\phi}\mathbf{e}^{a}\right) & = & e^{\phi}\mathbf{e}^{b}\wedge\tilde{\boldsymbol{\omega}}_{\;b}^{a}+\left(\boldsymbol{\omega}+\mathbf{d}\phi\right)\wedge e^{\phi}\mathbf{e}^{a}\\
e^{\phi}\mathbf{d}\phi\wedge\mathbf{e}^{a}+e^{\phi}\mathbf{d}\mathbf{e}^{a} & = & e^{\phi}\mathbf{e}^{b}\wedge\tilde{\boldsymbol{\omega}}_{\;b}^{a}+\left(\boldsymbol{\omega}+\mathbf{d}\phi\right)\wedge e^{\phi}\mathbf{e}^{a}\\
\mathbf{e}^{b}\wedge\boldsymbol{\omega}_{\;b}^{a} & = & \mathbf{e}^{b}\wedge\tilde{\boldsymbol{\omega}}_{\;b}^{a}
\end{eqnarray*}
and therefore $\tilde{\boldsymbol{\omega}}_{\;b}^{a}=\boldsymbol{\omega}_{\;b}^{a}$.
It follows from eq.(\ref{Structure equation for Weyl spin connection})
that the full Weyl curvature tensor is scale invariant, $\tilde{\boldsymbol{\mathfrak{R}}}_{\;b}^{a}=\boldsymbol{\mathfrak{R}}_{\;b}^{a}$.

Substituting into the curvature, the algebra is identical to that
leading up to eq.(\ref{Change in curvature}), with $\phi_{a}$ replaced
by $-W_{a}$. This results in
\begin{eqnarray*}
\boldsymbol{\mathfrak{R}}_{\;b}^{a} & = & \mathbf{R}_{\;b}^{a}-2\Delta_{db}^{ac}\left(W_{c;e}+W_{e}W_{c}-\frac{1}{2}W^{2}\eta_{ce}\right)\mathbf{e}^{e}\wedge\mathbf{e}^{d}
\end{eqnarray*}
which decomposes into three parts when we separate the symmetric and
antisymmetric parts of the trace term. With
\begin{eqnarray*}
\boldsymbol{\Omega} & = & \mathbf{d}\boldsymbol{\omega}\\
\Omega_{ab} & = & W_{\left[b;a\right]}
\end{eqnarray*}
we have
\begin{eqnarray}
\boldsymbol{\mathfrak{R}}_{\;b}^{a} & = & \mathbf{C}_{\;b}^{a}-2\Delta_{db}^{ac}\left(\mathcal{R}_{ce}+W_{\left(c;e\right)}+W_{e}W_{c}-\frac{1}{2}W^{2}\eta_{ce}\right)\mathbf{e}^{e}\wedge\mathbf{e}^{d}-2\Delta_{db}^{ac}\Omega_{ec}\mathbf{e}^{e}\wedge\mathbf{e}^{d}\label{Weyl curvature tensor}
\end{eqnarray}
where $\mathbf{R}_{\;b}^{a}=\mathbf{d}\boldsymbol{\alpha}_{\;b}^{a}-\boldsymbol{\alpha}_{\;b}^{c}\wedge\boldsymbol{\alpha}_{\;c}^{a}$
is the Riemannian part of the curvature. Carrying out the decomposition
of the curvature into trace and trace-free parts, we find the relationship
between the Weyl and Schouten tensors of the Weyl and Riemannian geometries.
In addition, the asymmetry of the Ricci tensor gives rise to a third
independent component, the dilatational curvature:
\begin{eqnarray*}
\boldsymbol{\mathfrak{C}}_{\;b}^{a} & = & \mathbf{C}_{\;b}^{a}\\
\boldsymbol{\mathscr{R}}_{a} & = & \boldsymbol{\mathcal{R}}_{a}+\left(W_{\left(a;c\right)}-W_{a}W_{c}+\frac{1}{2}W^{2}\eta_{ac}\right)\wedge\mathbf{e}^{c}\\
\boldsymbol{\Omega} & = & W_{\left[b;a\right]}\mathbf{e}^{a}\wedge\mathbf{e}^{b}
\end{eqnarray*}
or in components,
\begin{equation}
\mathfrak{\mathscr{R}}_{ab}=\mathcal{R}_{ab}+W_{\left(a;b\right)}-W_{a}W_{b}+\frac{1}{2}W^{2}\eta_{ab}\label{Weyl-Schouten tensor}
\end{equation}
We define the Weyl-Schouten tensor $\mathfrak{\mathscr{R}}_{ab}$
to be this symmetric part only.

In an integrable Weyl geometry, defined as one in which the dilatational
curvature, $\boldsymbol{\Omega}$, vanishes, there exists a conformal
transformation which makes the Weyl vector vanish, $W_{a}=0$. In
this gauge, the Weyl-Schouten tensor reduces to the Schouten tensor.
The gravitational field, $\mathbf{C}_{\;b}^{a}$, is the same in all
gauges.

The modification of eq.(\ref{Weyl curvature tensor}) in the presence
of torsion follows immediately since it only changes the Weyl connection
of eq.(\ref{Weyl connection}) by the contorsion tensor,
\[
\hat{\boldsymbol{\omega}}_{\;b}^{a}=\boldsymbol{\alpha}_{\;b}^{a}-2\Delta_{db}^{ac}W_{c}\mathbf{e}^{d}-C_{\;\;bc}^{a}\mathbf{e}^{c}=\boldsymbol{\omega}_{\;b}^{a}-\mathbf{C}_{\;\;b}^{a}
\]
This changes the curvature to
\begin{eqnarray*}
\hat{\boldsymbol{\mathfrak{R}}}_{\;b}^{a} & = & \boldsymbol{\mathfrak{R}}_{\;b}^{a}-\mathbf{D}\mathbf{C}_{\;\;b}^{a}+\mathbf{C}_{\;\;b}^{c}\land\mathbf{C}_{\;\;c}^{a}
\end{eqnarray*}

\section{Scale invariant gravity}

We now turn to the formulation of a scale invariant gravity theory,
based in a Weyl geometry. For this we must construct a Lorentz- and
dilatation-invariant action functional from the curvature and any
other available tensors. As we have noted, the conformal weight of
the curvature components in an orthonormal basis is $-2$. Since $g_{\mu\nu}=e_{\mu}^{\;a}e_{\nu}^{\;b}\eta_{ab}$,
the Minkowski metric has conformal weight zero, making the conformal
weight of the Weyl-Ricci scalar equal to $-2$ as expected. This introduces
a difficulty in writing a scale invariant action in dimensions greater
than two, since the volume element has weight $+n$ in $n$-dimensions.

\subsection{Actions nonlinear in the curvature}

In $2n$-dimensions, we may use $n$-products of the curvature:
\[
S=\intop\boldsymbol{\mathfrak{R}}^{ab}\wedge\boldsymbol{\mathfrak{R}}^{cd}\wedge\cdots\wedge\boldsymbol{\mathfrak{R}}^{ef}Q_{abcd\cdots ef}
\]
where $Q_{abcd\cdots ef}$ is a rank-$n$ invariant tensor. In $4$-dim
the most general such action is curvature-quadratic action, and takes
the form

\[
S=\intop\left(\alpha\mathfrak{R}^{abcd}\mathfrak{R}_{abcd}+\beta\mathfrak{R}^{ab}\mathfrak{R}_{ab}+\gamma\mathfrak{R}^{2}\right)\sqrt{-g}d^{4}x
\]
However, this may be simplified using the invariance of the Euler
character. Variation of the Gauss-Bonnet combination for the Euler
character $\chi=-\frac{1}{32\pi^{2}}\intop\boldsymbol{\mathfrak{R}}^{ab}\wedge\boldsymbol{\mathfrak{R}}^{cd}\varepsilon_{abcd}$,
gives
\begin{eqnarray*}
\delta\chi & = & \delta\intop\boldsymbol{\mathfrak{R}}^{ab}\wedge\boldsymbol{\mathfrak{R}}^{cd}\varepsilon_{abcd}\\
 & = & 2\intop\left(\mathbf{d}\left(\delta\boldsymbol{\omega}^{ab}\right)-\left(\delta\boldsymbol{\omega}^{eb}\right)\wedge\boldsymbol{\omega}_{\;e}^{a}-\left(\delta\boldsymbol{\omega}^{ae}\right)\wedge\boldsymbol{\omega}_{\;e}^{b}\right)\boldsymbol{\mathfrak{R}}^{cd}\varepsilon_{abcd}\\
 & = & 2\intop\mathbf{D}\left(\delta\boldsymbol{\omega}^{ab}\right)\wedge\boldsymbol{\mathfrak{R}}^{cd}\varepsilon_{abcd}\\
 & = & 2\intop\left(\mathbf{D}\left(\delta\boldsymbol{\omega}^{ab}\wedge\boldsymbol{\mathfrak{R}}^{cd}\varepsilon_{abcd}\right)+\delta\boldsymbol{\omega}^{ab}\wedge\mathbf{D}\boldsymbol{\mathfrak{R}}^{cd}\varepsilon_{abcd}\right)
\end{eqnarray*}
and this vanishes identically when we use the second Bianchi identity,
$\mathbf{D}\boldsymbol{\mathfrak{R}}^{cd}\equiv0$, and let the variation
vanish on the boundary,
\begin{eqnarray*}
\delta\chi & = & 2\intop_{V}\mathbf{D}\left(\delta\boldsymbol{\omega}^{ab}\wedge\boldsymbol{\mathfrak{R}}^{cd}\varepsilon_{abcd}\right)\\
 & = & 2\intop_{V}\mathbf{d}\left(\delta\boldsymbol{\omega}^{ab}\wedge\boldsymbol{\mathfrak{R}}^{cd}\varepsilon_{abcd}\right)\\
 & = & 2\left.\left(\delta\boldsymbol{\omega}^{ab}\wedge\boldsymbol{\mathfrak{R}}^{cd}\varepsilon_{abcd}\right)\right|_{\delta V}\\
 & = & 0
\end{eqnarray*}
The additon of any multiple of the Euler character to the action therefore
makes no contribution to the field equations.

We expand the Euler character as follows. Define a convenient volume
element as the dual of unity, $\boldsymbol{\Phi}\equiv\,^{*}1$. Then:
\begin{eqnarray*}
\boldsymbol{\Phi} & \equiv & \,^{*}1\\
 & = & \frac{1}{4!}\varepsilon_{abcd}\mathbf{e}^{a}\wedge\mathbf{e}^{b}\wedge\mathbf{e}^{c}\wedge\mathbf{e}^{d}\\
^{*}\boldsymbol{\Phi} & = & \,^{*}\left(\frac{1}{4!}\varepsilon_{abcd}\mathbf{e}^{a}\wedge\mathbf{e}^{b}\wedge\mathbf{e}^{c}\wedge\mathbf{e}^{d}\right)\\
 & = & \frac{1}{4!}\varepsilon_{abcd}\varepsilon^{abcd}\\
 & = & -1
\end{eqnarray*}
In a coordinate basis, $\boldsymbol{\Phi}=\frac{1}{4!}\sqrt{-g}\varepsilon_{\mu\nu\alpha\beta}\mathbf{d}x^{\mu}\wedge\mathbf{d}x^{\nu}\wedge\mathbf{d}x^{\alpha}\wedge\mathbf{d}x^{\beta}$,
and if we ignore orientation this is simply $\sqrt{-g}d^{4}x$. From
the definition of $\mathbf{\Phi}$ it follows that
\begin{eqnarray*}
\varepsilon^{abcd}\boldsymbol{\Phi} & = & \frac{1}{4!}\varepsilon^{abcd}\varepsilon_{efgh}\mathbf{e}^{e}\wedge\mathbf{e}^{f}\wedge\mathbf{e}^{g}\wedge\mathbf{e}^{h}\\
 & = & \frac{1}{4!}\left(-4!\delta_{efgh}^{abcd}\right)\mathbf{e}^{e}\wedge\mathbf{e}^{f}\wedge\mathbf{e}^{g}\wedge\mathbf{e}^{h}\\
 & = & -\mathbf{e}^{a}\wedge\mathbf{e}^{b}\wedge\mathbf{e}^{c}\wedge\mathbf{e}^{c}
\end{eqnarray*}
where\footnote{We check the normalization by contracting all pairs of indices, $\left(ae\right),\left(bf\right),\left(cg\right),\left(dh\right)$:
\begin{eqnarray*}
4! & = & \delta_{a}^{a}\left(\delta_{b}^{b}\left(\delta_{c}^{c}\delta_{d}^{d}-\delta_{c}^{d}\delta_{d}^{c}\right)+\delta_{b}^{c}\left(\delta_{c}^{d}\delta_{d}^{b}-\delta_{c}^{b}\delta_{d}^{d}\right)+\delta_{b}^{d}\left(\delta_{c}^{b}\delta_{d}^{c}-\delta_{c}^{c}\delta_{d}^{b}\right)\right)\\
 &  & -\delta_{a}^{b}\left(\delta_{b}^{a}\left(\delta_{c}^{c}\delta_{d}^{d}-\delta_{c}^{d}\delta_{d}^{c}\right)+\delta_{b}^{c}\left(\delta_{c}^{d}\delta_{d}^{a}-\delta_{c}^{e}\delta_{d}^{d}\right)+\delta_{b}^{d}\left(\delta_{c}^{a}\delta_{d}^{c}-\delta_{c}^{c}\delta_{d}^{a}\right)\right)\\
 &  & -\delta_{a}^{b}\left(\delta_{b}^{a}\left(\delta_{c}^{c}\delta_{d}^{d}-\delta_{c}^{d}\delta_{d}^{c}\right)+\delta_{b}^{c}\left(\delta_{c}^{d}\delta_{d}^{a}-\delta_{c}^{e}\delta_{d}^{d}\right)+\delta_{b}^{d}\left(\delta_{c}^{a}\delta_{d}^{c}-\delta_{c}^{c}\delta_{d}^{a}\right)\right)\\
 &  & -\delta_{a}^{c}\left(\delta_{b}^{b}\left(\delta_{c}^{a}\delta_{d}^{d}-\delta_{c}^{d}\delta_{d}^{a}\right)+\delta_{b}^{a}\left(\delta_{c}^{d}\delta_{d}^{b}-\delta_{c}^{f}\delta_{d}^{d}\right)+\delta_{b}^{d}\left(\delta_{c}^{b}\delta_{d}^{a}-\delta_{c}^{a}\delta_{d}^{b}\right)\right)\\
 &  & -\delta_{a}^{d}\left(\delta_{b}^{b}\left(\delta_{c}^{c}\delta_{d}^{a}-\delta_{c}^{a}\delta_{d}^{c}\right)+\delta_{b}^{c}\left(\delta_{c}^{a}\delta_{d}^{b}-\delta_{c}^{b}\delta_{d}^{a}\right)+\delta_{b}^{a}\left(\delta_{c}^{b}\delta_{d}^{c}-\delta_{c}^{c}\delta_{d}^{b}\right)\right)\\
 & = & 4\left(4\left(16-4\right)+4-16+4-16\right)-\left(64-16+4-16+4-16\right)\\
 &  & -\left(64-16+4-16+4-16\right)-\left(64-16+4-16+4-16\right)\\
 & = & 24
\end{eqnarray*}
} $\varepsilon^{abcd}\varepsilon_{efgh}=-4!\delta_{efgh}^{abcd}$.
Therefore, substituting $\mathbf{e}^{a}\wedge\mathbf{e}^{b}\wedge\mathbf{e}^{c}\wedge\mathbf{e}^{d}=-\varepsilon^{abcd}\boldsymbol{\Phi}$
into the expression for the Euler character,
\begin{eqnarray*}
\chi & = & -\frac{1}{32\pi^{2}}\intop\boldsymbol{\mathfrak{R}}^{ab}\wedge\boldsymbol{\mathfrak{R}}^{cd}\varepsilon_{abcd}\\
 & = & -\frac{1}{128\pi^{2}}\intop\mathfrak{R}_{\quad ef}^{ab}\mathfrak{R}_{\quad gh}^{cd}\mathbf{e}^{e}\wedge\mathbf{e}^{f}\wedge\mathbf{e}^{g}\wedge\mathbf{e}^{h}\varepsilon_{abcd}\\
 & = & \frac{1}{128\pi^{2}}\intop\mathfrak{R}_{\quad ef}^{ab}\mathfrak{R}_{\quad gh}^{cd}\varepsilon^{efgh}\varepsilon_{abcd}\boldsymbol{\Phi}\\
 & = & \frac{1}{128\pi^{2}}\intop\mathfrak{R}_{\quad ef}^{ab}\mathfrak{R}_{\quad gh}^{cd}4!\delta_{abcd}^{efgh}\boldsymbol{\Phi}
\end{eqnarray*}
Expanding the antisymmetric $\delta_{abcd}^{efgh}$,
\begin{eqnarray*}
4!\delta_{abcd}^{efgh} & \equiv & \delta_{a}^{e}\left(\delta_{b}^{f}\left(\delta_{c}^{g}\delta_{d}^{h}-\delta_{c}^{h}\delta_{d}^{g}\right)+\delta_{b}^{g}\left(\delta_{c}^{h}\delta_{d}^{f}-\delta_{c}^{f}\delta_{d}^{h}\right)+\delta_{b}^{h}\left(\delta_{c}^{f}\delta_{d}^{g}-\delta_{c}^{g}\delta_{d}^{f}\right)\right)\\
 &  & -\delta_{a}^{f}\left(\delta_{b}^{e}\left(\delta_{c}^{g}\delta_{d}^{h}-\delta_{c}^{h}\delta_{d}^{g}\right)+\delta_{b}^{g}\left(\delta_{c}^{h}\delta_{d}^{e}-\delta_{c}^{e}\delta_{d}^{h}\right)+\delta_{b}^{h}\left(\delta_{c}^{e}\delta_{d}^{g}-\delta_{c}^{g}\delta_{d}^{e}\right)\right)\\
 &  & -\delta_{a}^{g}\left(\delta_{b}^{f}\left(\delta_{c}^{e}\delta_{d}^{h}-\delta_{c}^{h}\delta_{d}^{e}\right)+\delta_{b}^{e}\left(\delta_{c}^{h}\delta_{d}^{f}-\delta_{c}^{f}\delta_{d}^{h}\right)+\delta_{b}^{h}\left(\delta_{c}^{f}\delta_{d}^{e}-\delta_{c}^{e}\delta_{d}^{f}\right)\right)\\
 &  & -\delta_{a}^{h}\left(\delta_{b}^{f}\left(\delta_{c}^{g}\delta_{d}^{e}-\delta_{c}^{e}\delta_{d}^{g}\right)+\delta_{b}^{g}\left(\delta_{c}^{e}\delta_{d}^{f}-\delta_{c}^{f}\delta_{d}^{e}\right)+\delta_{b}^{e}\left(\delta_{c}^{f}\delta_{d}^{g}-\delta_{c}^{g}\delta_{d}^{f}\right)\right)
\end{eqnarray*}
we explicitly write out the integrand of the Euler character in the
Gauss-Bonnet form\footnote{
\begin{eqnarray*}
4!\mathfrak{R}_{\quad ef}^{ab}\mathfrak{R}_{\quad gh}^{cd}\delta_{abcd}^{efgh} & = & \mathfrak{R}_{\quad ef}^{ab}\mathfrak{R}_{\quad gh}^{cd}\left(\delta_{a}^{e}\left(\delta_{b}^{f}\left(\delta_{c}^{g}\delta_{d}^{h}-\delta_{c}^{h}\delta_{d}^{g}\right)+\delta_{b}^{g}\left(\delta_{c}^{h}\delta_{d}^{f}-\delta_{c}^{f}\delta_{d}^{h}\right)+\delta_{b}^{h}\left(\delta_{c}^{f}\delta_{d}^{g}-\delta_{c}^{g}\delta_{d}^{f}\right)\right)\right)\\
 &  & -\mathfrak{R}_{\quad ef}^{ab}\mathfrak{R}_{\quad gh}^{cd}\delta_{a}^{f}\left(\delta_{b}^{e}\left(\delta_{c}^{g}\delta_{d}^{h}-\delta_{c}^{h}\delta_{d}^{g}\right)+\delta_{b}^{g}\left(\delta_{c}^{h}\delta_{d}^{e}-\delta_{c}^{e}\delta_{d}^{h}\right)+\delta_{b}^{h}\left(\delta_{c}^{e}\delta_{d}^{g}-\delta_{c}^{g}\delta_{d}^{e}\right)\right)\\
 &  & -\mathfrak{R}_{\quad ef}^{ab}\mathfrak{R}_{\quad gh}^{cd}\delta_{a}^{g}\left(\delta_{b}^{f}\left(\delta_{c}^{e}\delta_{d}^{h}-\delta_{c}^{h}\delta_{d}^{e}\right)+\delta_{b}^{e}\left(\delta_{c}^{h}\delta_{d}^{f}-\delta_{c}^{f}\delta_{d}^{h}\right)+\delta_{b}^{h}\left(\delta_{c}^{f}\delta_{d}^{e}-\delta_{c}^{e}\delta_{d}^{f}\right)\right)\\
 &  & -\mathfrak{R}_{\quad ef}^{ab}\mathfrak{R}_{\quad gh}^{cd}\delta_{a}^{h}\left(\delta_{b}^{f}\left(\delta_{c}^{g}\delta_{d}^{e}-\delta_{c}^{e}\delta_{d}^{g}\right)+\delta_{b}^{g}\left(\delta_{c}^{e}\delta_{d}^{f}-\delta_{c}^{f}\delta_{d}^{e}\right)+\delta_{b}^{e}\left(\delta_{c}^{f}\delta_{d}^{g}-\delta_{c}^{g}\delta_{d}^{f}\right)\right)\\
 & = & 2\left(\mathfrak{R}_{\quad ab}^{ab}\mathfrak{R}_{\quad cd}^{cd}+\mathfrak{R}_{\quad ad}^{ab}\mathfrak{R}_{\quad bc}^{cd}+\mathfrak{R}_{\quad ac}^{ab}\mathfrak{R}_{\quad bd}^{cd}\right)-2\left(\mathfrak{R}_{\quad ba}^{ab}\mathfrak{R}_{\quad cd}^{cd}+\mathfrak{R}_{\quad da}^{ab}\mathfrak{R}_{\quad bc}^{cd}+\mathfrak{R}_{\quad ca}^{ab}\mathfrak{R}_{\quad db}^{cd}\right)\\
 &  & -2\left(\mathfrak{R}_{\quad cb}^{ab}\mathfrak{R}_{\quad ad}^{cd}+\mathfrak{R}_{\quad bd}^{ab}\mathfrak{R}_{\quad ac}^{cd}+\mathfrak{R}_{\quad dc}^{ab}\mathfrak{R}_{\quad ab}^{cd}\right)-2\left(\mathfrak{R}_{\quad db}^{ab}\mathfrak{R}_{\quad ca}^{cd}+\mathfrak{R}_{\quad bc}^{ab}\mathfrak{R}_{\quad da}^{cd}+\mathfrak{R}_{\quad cd}^{ab}\mathfrak{R}_{\quad ba}^{cd}\right)\\
 & = & 4\left(\mathfrak{R}^{2}-4\mathfrak{R}_{\;d}^{b}\mathfrak{R}_{\;b}^{d}+\mathfrak{R}^{abcd}\mathfrak{R}_{abcd}\right)
\end{eqnarray*}
},
\begin{eqnarray*}
\boldsymbol{\mathfrak{R}}^{ab}\wedge\boldsymbol{\mathfrak{R}}^{cd}\varepsilon_{abcd} & = & -\frac{1}{4}\mathfrak{R}_{\quad ef}^{ab}\mathfrak{R}_{\quad gh}^{cd}4!\delta_{abcd}^{efgh}\boldsymbol{\Phi}\\
 & = & -\left(\mathfrak{R}^{2}-4\mathfrak{R}_{\;d}^{b}\mathfrak{R}_{\;b}^{d}+\mathfrak{R}^{abcd}\mathfrak{R}_{abcd}\right)
\end{eqnarray*}
so that
\begin{eqnarray*}
\chi & = & \frac{1}{32\pi^{2}}\intop\left(\mathfrak{R}^{2}-4\mathfrak{R}_{\;d}^{b}\mathfrak{R}_{\;b}^{d}+\mathfrak{R}^{abcd}\mathfrak{R}_{abcd}\right)\boldsymbol{\Phi}
\end{eqnarray*}
The invariance of $\chi$ allows us to replace
\[
\intop\mathfrak{R}^{abcd}\mathfrak{R}_{abcd}\boldsymbol{\Phi}=32\pi^{2}\chi-\intop\left(\mathfrak{R}^{2}-4\mathfrak{R}_{\;d}^{b}\mathfrak{R}_{\;b}^{d}\right)\boldsymbol{\Phi}
\]
Dropping the invariant first term leaves the most general curvature-quadratic
action in the form
\begin{equation}
S=\intop\left(a\mathfrak{R}^{2}+b\mathfrak{R}^{ab}\mathfrak{R}_{ab}\right)\sqrt{-g}d^{4}x\label{Action for quadratic gravity}
\end{equation}
for constants $a,b$. Quadratic gravity theories, especially the $b=0$
case, have often been studied because the scale invariance allows
the theory to be renormalizable. However, fourth order field equations
such as those resulting from eq.(\ref{Action for quadratic gravity})
are sometimes found to introduce ghosts in the quantum theory. The
Einstein-Hilbert term may be included as well, but while this has
desirable effects, it breaks the scale invariance we examine here.
For further discussion and references on quadratic gravity, see \cite{Alvarez--Gaume et al}. 

The particular case of Weyl (conformal) gravity deserves mention.
As first shown by Bach \cite{Bach}, the \emph{fully} conformally
action for Weyl gravity may be written as
\begin{eqnarray}
S_{W} & = & \alpha\intop C^{abcd}C_{abcd}\sqrt{-g}d^{4}x\nonumber \\
 & = & \alpha\intop\left(R^{abcd}R_{abcd}-2R^{bd}R_{bd}+\frac{1}{3}R^{2}\right)\sqrt{-g}d^{4}x\nonumber \\
 & = & \left(32\pi^{2}\alpha\chi-\alpha\intop\left(R^{2}-4R_{\;d}^{b}R_{\;b}^{d}\right)\sqrt{-g}d^{4}x\right)+\alpha\intop\left(-2R^{bd}R_{bd}+\frac{1}{3}R^{2}\right)\sqrt{-g}d^{4}x\nonumber \\
 & = & 32\pi^{2}\alpha\chi+2\alpha\intop\left(R_{\;d}^{b}R_{\;b}^{d}-\frac{1}{3}R^{2}\right)\sqrt{-g}d^{4}x\label{Action for Weyl gravity}
\end{eqnarray}
Naturally, fourth order field equations result from \emph{metric}
variation of eq.(\ref{Action for Weyl gravity}). However, it is shown
in \cite{Wheeler 2014} that varying the full conformal connection
in the action \ref{Action for Weyl gravity} gives the additional
integrability condition to reduce the fourth order equations to the
Einstein equation. 

Quadratic gravity applies only in four dimensions, with dimension
$2n$ theories having correspondingly more complicated field equations.
Instead, we consider Weyl invariant theories linear in the curvature. 

\subsection{Actions linear in the curvature}

There are two classes of gravitational theories with Lorentz and dilatational
symmetry with actions linear in the curvatures. Both may be written
in any dimension. 

\subsubsection{Biconformal gravity}

Based in geometries first developed by Ivanov and Niederle \cite{Ivanov Niederle I 1982,Ivanov Niederle II 1982}
and Wheeler \cite{Wheeler JMP}, what is now called biconformal gravity
takes place in a $2n$-dimensional symplectic manifold. The canonical
conjucacy of ths space makes the volume element dimensionless, allowing
for a curvature-linear action in any dimension \cite{WW} of the form
\begin{equation}
S=\int(\alpha\boldsymbol{\mathbf{\Omega}}_{\;\;b_{1}}^{a_{1}}+\beta\delta_{b_{1}}^{a_{1}}\boldsymbol{\mathbf{\Omega}}+\gamma\mathbf{e}^{a_{1}}\land\mathbf{f}_{b_{1}})\land\mathbf{e}^{a_{2}}\land\cdots\land\mathbf{e}^{a_{n}}\land\mathbf{f}_{b_{1}}\land\cdots\land\mathbf{f}_{b_{1}}\varepsilon^{b_{1}...b_{n}}\varepsilon_{a_{1}...a_{n}}\label{Biconformal action}
\end{equation}
The resulting theory describes $n$-dimensional gravity on an $n$-dimensional
Lagrangian submanifold. Arising as the quotient of the conformal group
by the Weyl group, the theory leads to scale invariant general relativity.
Because the underlying structure is the full conformal group instead
of the Weyl group, resulting in Kähler geometry instead of Weyl geometry,
we will not go into further detail here.

\subsubsection{The Dirac theory}

An alternative approach to scale invariant gravity was developed by
Dirac in an attempt to give rigor to his Large Numbers Hypothesis
\cite{Dirac LNH}, the idea that extremely large dimensionless numbers
in the description of nature should be related to one another. In
\cite{Dirac Large Numbers}, Dirac presents a scale invariant gravity
theory in which the gravitational constant varies with time in such
a way that the large dimensionless magnitude constructed from the
fundamental charge $e$ and $G$ is related to the age of the universe.
The result follows from a single, simple solution to the scale invariant
theory. Here we examine the scale invariant theory without further
discussion of the Large Numbers Hypothesis.

In the Dirac theory, scale invariance is achieved by including a gravitationally
coupled scalar field in addition to the curvature. There is a wide
literature on scalar fields coupled to gravity. Fierz \cite{Fierz}
and Jordan \cite{Jordan} showed that the energy-momentum tensor of
scalar theories may sometimes be unphysical. This flaw is corrected
by the Brans-Dicke scalar-tensor theory \cite{Brans Dicke}, and discussion
contiues. In more recent work, \cite{Romero et al}, Romero, Fonseca-Neto,
and Pucheu study the relationship between Brans-Dicke theory and integrable
Weyl geometry is studied. See also the history by Brans \cite{Brans}.

Dirac begins with an action which in our notation takes the form
\begin{equation}
S_{D}=\intop\left(\frac{1}{4}\Omega^{\mu\nu}\Omega_{\mu\nu}-\varphi^{2}R+6g^{\mu\nu}D_{\mu}\varphi D_{\nu}\varphi+c\varphi^{4}\right)\sqrt{-g}d^{n}x\label{Original Dirac action}
\end{equation}
where $\Omega_{\mu\nu}=W_{\mu,\nu}-W_{\mu,\nu}$ is the dilatational
curvature. Dropping the dilatation, and allowing an arbitrary constant
multiplying the kinetic term gives the Brans-Dicke scalar-tensor theory,
usually written as
\[
S_{D}=\intop\left(\varphi R+\frac{\omega_{0}}{\varphi}g^{\mu\nu}D_{\mu}\varphi D_{\nu}\varphi+\mathcal{L}_{m}\right)\sqrt{-g}d^{n}x
\]
where $\mathcal{L}_{m}$ is a matter Lagrangian.

In the next Section, we take a similar but slightly different approach,
allowing torsion and a mass term but no quadratic dilatational term.
Our treatment also differs by our use of a Palatini variation, so
the metric and connection are regarded as independent variables. Finally,
we consider arbitrary spacetime dimension. We find a locally scale
invariant theory that exactly reproduces general relativity as soon
as a suitable definition of the unit of length is made.

\section{Curvature linear, scale invariant gravity in any dimension}

Beginning with the action for a Klein-Gordon scalar field, $\varphi$,
in curved, $n$-dimensional Weyl geometry, we add a non-quadratic
term analogous to Dirac's $\varphi^{4}$ potential,
\begin{equation}
S_{\varphi}=\intop\left(g^{\mu\nu}D_{\mu}\varphi D_{\nu}\varphi+\frac{m^{2}c^{2}}{\hbar^{2}}\varphi^{2}+\beta\varphi^{s}\right)\sqrt{-g}d^{n}x\label{Klein-Gordon action}
\end{equation}
where the covariant derivative of $\varphi$ is $D_{\mu}\varphi=\partial_{\mu}\varphi-\lambda W_{\mu}$,
we include a gravitational term of the formwhere $\boldsymbol{\Phi}\equiv\,^{*}1=\frac{1}{n!}\varepsilon_{a\cdots b}\mathbf{e}^{a}\wedge\cdots\wedge\mathbf{e}^{b}$
in $n$-dim and the power $k$ will be chosen to make the full action
scale invariant. There is no scalar we can form which is linear in
the dilatational curvature. We choose units $\hbar=c=1$. 

To include gravity, we multiply a power of the scalar field times
the scalar curvature. Thus, we arrive at
\begin{equation}
S=\intop\left(\alpha\varphi^{k}\mathfrak{R}+g^{\mu\nu}D_{\mu}\varphi D_{\nu}\varphi+m^{2}\varphi^{2}+\beta\varphi^{s}\right)\sqrt{-g}d^{n}x\label{Diraclike action}
\end{equation}
If the scalar field, metric, curvature and geometric mass scale as
\begin{eqnarray*}
\varphi & \rightarrow & e^{\lambda\phi}\varphi\\
g_{\mu\nu} & \rightarrow & e^{2\phi}g_{\mu\nu}\\
g & \rightarrow & e^{2n\phi}g\\
\mathfrak{R} & \rightarrow & e^{-2\phi}\mathfrak{R}\\
m^{2}=\frac{m^{2}c^{2}}{\hbar^{2}} & \rightarrow & e^{-2\phi}m^{2}\\
\alpha,\beta &  & dimensionless
\end{eqnarray*}
then $S$ scales as
\begin{eqnarray*}
\tilde{S} & = & \intop\left(\alpha\tilde{\varphi}^{k}\tilde{\mathfrak{R}}+\tilde{g}^{\mu\nu}D_{\mu}\tilde{\varphi}D_{\nu}\tilde{\varphi}+m^{2}\tilde{\varphi}^{2}+\beta\tilde{\varphi}^{s}\right)\sqrt{-\tilde{g}}d^{n}x\\
 & = & \intop\left(\alpha e^{\left(\lambda k-2\right)\phi}\varphi^{k}\mathfrak{R}+e^{-2\phi}e^{2\lambda\phi}g^{\mu\nu}D_{\mu}\varphi D_{\nu}\varphi+e^{-2\phi}e^{2\lambda\phi}m^{2}\varphi^{2}+e^{s\lambda\phi}\beta\varphi^{s}\right)e^{n\phi}\sqrt{-g}d^{n}x\\
 & = & \intop\left[e^{\left(\lambda k-2+n\right)\phi}\left(\alpha\varphi^{k}\mathfrak{R}\right)+e^{\left(n-2+2\lambda\right)\phi}\left(g^{\mu\nu}D_{\mu}\varphi D_{\nu}\varphi+m^{2}\varphi^{2}\right)+e^{\left(n+s\lambda\right)\phi}\beta\varphi^{s}\right]\sqrt{-g}d^{n}x
\end{eqnarray*}
and is therefore locally scale invariant if
\begin{eqnarray*}
\lambda & = & -\frac{n-2}{2}\\
k & = & 2\\
s & = & \frac{2n}{n-2}
\end{eqnarray*}
We may make these assignments in any dimension greater than two, resulting
in
\begin{equation}
S=\intop\left(\alpha\varphi^{2}\mathfrak{R}+g^{\mu\nu}D_{\mu}\varphi D_{\nu}\varphi+m^{2}\varphi^{2}+\beta\varphi^{\frac{2n}{n-2}}\right)\sqrt{-g}d^{n}x\label{Scale invariant Dirac action}
\end{equation}

\subsection{Variation of the action}

We consider the Palatini variation of $S$, varying the solder form,
spin connection, Weyl vector and scalar field, $\left(\mathbf{e}^{a},\,\boldsymbol{\omega}_{\;\;b}^{a},\,W_{a},\,\varphi\right)$,
independently. The variation of the solder form is most easily accomplished
by varying the metric. These are equivalent, since
\begin{eqnarray*}
\delta g_{\alpha\beta} & = & 2\eta_{ab}e_{\alpha}^{\;\;a}\delta e_{\beta}^{\;\;b}
\end{eqnarray*}
and conversely
\begin{eqnarray*}
\delta e_{\alpha}^{\;\;a} & = & 2\eta^{ab}e_{b}^{\;\;\beta}\delta g_{\alpha\beta}
\end{eqnarray*}
The connection variation is easiest using differential forms and varying
$\boldsymbol{\omega}_{\;\;b}^{a}$ directly.

\subsubsection{Metric variation}

Writing $\mathfrak{R}=g^{\mu\nu}\mathfrak{R}_{\mu\nu}$ where $\mathfrak{R}_{\mu\nu}$
depends only on the spin connection, and noting that $\delta\sqrt{-g}=-\frac{1}{2}\sqrt{-g}g_{\mu\nu}\delta g^{\mu\nu}$,
the metric variation is

\[
0=\intop\delta g^{\mu\nu}\left[\alpha\varphi^{2}\mathfrak{R}_{\mu\nu}+D_{\mu}\varphi D_{\nu}\varphi-\frac{1}{2}g_{\mu\nu}\left(\alpha\varphi^{2}\mathfrak{R}+g^{\mu\nu}D_{\mu}\varphi D_{\nu}\varphi+m^{2}\varphi^{2}+\beta\varphi^{\frac{2n}{n-2}}\right)\right]\sqrt{-g}d^{n}x
\]
so we immediately get the field equation,
\begin{eqnarray*}
\alpha\varphi^{2}\left(\mathfrak{R}_{\mu\nu}-\frac{1}{2}g_{\mu\nu}\mathfrak{R}\right) & = & -D_{\mu}\varphi D_{\nu}\varphi+\frac{1}{2}g_{\mu\nu}\left(D^{\alpha}\varphi D_{\alpha}\varphi+m^{2}\varphi^{2}+\beta\varphi^{\frac{2n}{n-2}}\right)
\end{eqnarray*}
This takes the form of the scale covariant Einstein equation
\begin{eqnarray}
\mathfrak{R}_{\mu\nu}-\frac{1}{2}g_{\mu\nu}\mathfrak{R} & = & \frac{1}{\alpha}\left[-\frac{1}{\varphi^{2}}D_{\mu}\varphi D_{\nu}\varphi+\frac{1}{2}g_{\mu\nu}\left(\frac{1}{\varphi^{2}}D^{\alpha}\varphi D_{\alpha}\varphi+m^{2}+\beta\varphi^{\frac{4}{n-2}}\right)\right]\label{Einstein equation with sources}
\end{eqnarray}

\subsubsection{Scalar field variation}

The scalar field variation is
\begin{eqnarray*}
0 & = & \intop\left(2\alpha\delta\varphi\varphi\mathfrak{R}+2g^{\mu\nu}D_{\mu}\delta\varphi D_{\nu}\varphi+2m^{2}\varphi\delta\varphi+\frac{2n}{n-2}\beta\delta\varphi\varphi^{\frac{n+2}{n-2}}\right)\sqrt{-g}d^{n}x\\
 & = & \intop\delta\varphi\left(2\alpha\varphi\mathfrak{R}-2D_{\mu}\left(g^{\mu\nu}D_{\nu}\varphi\right)+2m^{2}\varphi+\frac{2n}{n-2}\beta\varphi^{\frac{n+2}{n-2}}\right)\sqrt{-g}d^{n}x
\end{eqnarray*}
and therefore we find a nonlinear wave equation coupled to the scalar
curvature, 
\begin{equation}
D^{a}D_{a}\varphi-m^{2}\varphi-\alpha\mathfrak{R}\varphi-\frac{n}{n-2}\beta\varphi^{\frac{n+2}{n-2}}=0\label{Wave equation for phi}
\end{equation}

\subsubsection{Weyl vector variation}

The Weyl vector only appears in the kinetic term for the scalar field,
where $D_{\mu}\varphi=\partial_{\mu}\varphi-\lambda W_{\mu}\varphi$.
Varying $W_{\mu}$,
\[
0=\intop\left(2g^{\mu\nu}\lambda\delta W_{\mu}\varphi D_{\nu}\varphi\right)\sqrt{-g}d^{n}x
\]
and therefore,
\begin{eqnarray}
D_{\alpha}\varphi & = & 0\label{Solution for Weyl vector}
\end{eqnarray}

This may immediately be solved for the Weyl vector
\begin{eqnarray*}
\mathbf{d}\varphi-\lambda\boldsymbol{\omega}\varphi & = & 0\\
\boldsymbol{\omega} & = & \mathbf{d}\left(\frac{1}{\lambda}\ln\varphi\right)\\
 & = & -\frac{2}{n-2}\frac{1}{\varphi}\mathbf{d}\varphi
\end{eqnarray*}
which implies an integrable Weyl geometry and the existence of a gauge
in which the Weyl vector vanishes. We easily find the gauge transformation
$\phi$ required to remove the Weyl vector. 
\begin{eqnarray*}
\tilde{\boldsymbol{\omega}} & = & \boldsymbol{\omega}+\mathbf{d}\phi\\
0 & = & \mathbf{d}\left(\frac{1}{\lambda}\ln\varphi\right)+\mathbf{d}\phi\\
\phi & = & -\frac{1}{\lambda}\ln\frac{\varphi}{\varphi_{0}}
\end{eqnarray*}
With this transformation, we have $\tilde{\boldsymbol{\omega}}=0$.
This now remains the case for arbitrary \emph{global }scale transformations.

The same transformation changes the scalar field according to
\begin{eqnarray*}
\tilde{\varphi} & = & \varphi e^{\lambda\phi}\\
 & = & \varphi e^{\lambda\left(-\frac{1}{\lambda}\ln\frac{\varphi}{\varphi_{0}}\right)}\\
 & = & \varphi_{0}
\end{eqnarray*}
so the transformation that removes the Weyl vector simultaneously
makes the scalar field constant. In an arbitrary gauge $\phi$, the
Weyl vector and scalar field become
\begin{eqnarray*}
W_{\mu} & = & \partial_{\mu}\phi\\
\varphi & = & \varphi_{0}e^{\lambda\phi}
\end{eqnarray*}
but the physical content of the theory remains the same.

If we were to allow curvature squared terms in the action, it would
involve $\boldsymbol{\Omega}^{*}\boldsymbol{\Omega}$ where $\Omega_{\mu\nu}=W_{\mu,\nu}-W_{\nu,\mu}$.
Such a dilatational curvature would lead, in general, to a nonintegrable
Weyl geometry. However, the physical constraints against such a geometry
are extremely strong \textendash{} we do not experience changes of
relative physical size.

A quadratic term, $\boldsymbol{\Omega}^{*}\boldsymbol{\Omega}$, is
a kinetic term for a nonintegrable Weyl vector. Except in dimension
$n=4$, scale invariance of such a kinetic term would also require
a factor of the scalar field,
\[
\frac{1}{4}\int\varphi^{\frac{2\left(n-4\right)}{n-2}}g^{\mu\alpha}g^{\nu\beta}\Omega_{\mu\nu}\Omega_{\alpha\beta}\sqrt{-g}d^{n}x
\]
This would lead to a field equation of the form
\[
\left(\varphi^{\frac{2\left(n-4\right)}{n-2}}\Omega^{\mu\nu}\right)_{;\nu}=2\lambda D^{\mu}\varphi
\]
which allows nontrivial $\varphi$but also a nontrivial dilatation,
$\Omega_{\mu\nu}$. The dilatational curvature would also act as a
source to the wave equation for $\varphi$,
\[
D^{\mu}D_{\mu}\varphi+m^{2}\varphi+\frac{n}{n-2}\beta\varphi^{\frac{n+2}{n-2}}+\alpha\varphi\mathfrak{R}+\frac{n-4}{n-2}\varphi^{\frac{n-6}{n-2}}\Omega^{\alpha\beta}\Omega_{\alpha\beta}=0
\]
and as an energy source for the Einstein equation in the form
\[
T_{\mu\nu}=\varphi^{\frac{2\left(n-4\right)}{n-2}}\left(\Omega_{\mu}^{\;\;\beta}\Omega_{\nu\beta}-\frac{1}{4}g_{\mu\nu}\Omega^{\alpha\beta}\Omega_{\alpha\beta}\right)
\]
We continue without the kinetic term, because of the unphysical nature
of dilatations.

\subsubsection{Connection variation}

The variation of the spin connection is much easier to work with than
the variation of the Christoffel connection. Since only the curvature
depends on the connection, we need only the first term in the action
and the equivalence
\begin{eqnarray*}
\frac{\alpha}{\left(n-2\right)!}\varphi^{2}\boldsymbol{\mathfrak{R}}^{ab}\wedge\mathbf{e}^{c}\wedge\cdots\wedge\mathbf{e}^{d}\varepsilon_{abc\cdots d} & = & \frac{\alpha}{\left(n-2\right)!}\varphi^{2}\frac{1}{2}\mathfrak{R}_{\quad ef}^{ab}\mathbf{e}^{e}\wedge\mathbf{e}^{f}\wedge\mathbf{e}^{c}\wedge\cdots\wedge\mathbf{e}^{d}\varepsilon_{abc\cdots d}\\
 & = & -\frac{\alpha}{\left(n-2\right)!}\varphi^{2}\frac{1}{2}\mathfrak{R}_{\quad ef}^{ab}\varepsilon^{efc\cdots d}\varepsilon_{abc\cdots d}\boldsymbol{\Phi}\\
 & = & \frac{\alpha}{\left(n-2\right)!}\varphi^{2}\frac{1}{2}\mathfrak{R}_{\quad ef}^{ab}\left(n-2\right)!\left(\delta_{a}^{e}\delta_{b}^{f}-\delta_{a}^{f}\delta_{b}^{e}\right)\boldsymbol{\Phi}\\
 & = & \alpha\varphi^{2}\mathfrak{R}\boldsymbol{\Phi}
\end{eqnarray*}

Therefore, replacing $\intop\alpha\varphi^{k}\mathfrak{R}\sqrt{-g}d^{n}x$
with $\intop\frac{\alpha}{\left(n-2\right)!}\varphi^{2}\delta_{\omega_{b}^{a}}\boldsymbol{\mathfrak{R}}^{ab}\wedge\mathbf{e}^{c}\wedge\cdots\wedge\mathbf{e}^{d}\varepsilon_{abc\cdots d}$,
we vary $\boldsymbol{\omega}_{\;\;b}^{a}$,
\begin{eqnarray*}
\delta_{\omega^{ab}}S & = & \intop\frac{\alpha}{\left(n-2\right)!}\varphi^{2}\delta_{\omega_{b}^{a}}\boldsymbol{\mathfrak{R}}^{ab}\wedge\mathbf{e}^{c}\wedge\cdots\wedge\mathbf{e}^{d}\varepsilon_{abc\cdots d}\\
 & = & \intop\frac{\alpha}{\left(n-2\right)!}\varphi^{2}\delta_{\omega^{ab}}\left(\mathbf{d}\boldsymbol{\omega}^{ab}-\eta_{ef}\boldsymbol{\omega}^{eb}\wedge\boldsymbol{\omega}^{af}\right)\wedge\mathbf{e}^{c}\wedge\cdots\wedge\mathbf{e}^{d}\varepsilon_{abc\cdots d}\\
 & = & \intop\frac{\alpha}{\left(n-2\right)!}\varphi^{2}\left(\mathbf{d}\delta\boldsymbol{\omega}^{ab}-\delta\boldsymbol{\omega}^{eb}\wedge\boldsymbol{\omega}_{\;\;e}^{a}-\delta\boldsymbol{\omega}^{ae}\wedge\boldsymbol{\omega}_{\;\;e}^{b}\right)\wedge\mathbf{e}^{c}\wedge\cdots\wedge\mathbf{e}^{d}\varepsilon_{abc\cdots d}\\
 & = & \intop\frac{\alpha}{\left(n-2\right)!}\varphi^{2}\mathbf{D}\left(\delta\boldsymbol{\omega}^{ab}\right)\wedge\mathbf{e}^{c}\wedge\cdots\wedge\mathbf{e}^{d}\varepsilon_{abc\cdots d}\\
 & = & \intop\frac{\alpha}{\left(n-2\right)!}\mathbf{D}\left(\varphi^{2}\left(\delta\boldsymbol{\omega}^{ab}\right)\wedge\mathbf{e}^{c}\wedge\cdots\wedge\mathbf{e}^{d}\varepsilon_{abc\cdots d}\right)-\intop\frac{\alpha}{\left(n-2\right)!}\left(2\varphi\left(\mathbf{D}\varphi\right)\delta\boldsymbol{\omega}_{\;\;b}^{a}\wedge\mathbf{e}^{c}\wedge\cdots\wedge\mathbf{e}^{d}\varepsilon_{abc\cdots d}\right)\\
 &  & +\intop\frac{\alpha}{\left(n-2\right)!}\left(\left(n-2\right)\varphi^{2}\delta\boldsymbol{\omega}^{ab}\wedge\mathbf{D}\mathbf{e}^{c}\wedge\cdots\wedge\mathbf{e}^{d}\varepsilon_{abc\cdots d}\right)
\end{eqnarray*}
The first term is a total divergence which we discard. Then, writing
$\delta\boldsymbol{\omega}^{ab}=\delta\omega_{\quad k}^{ab}\mathbf{e}^{k}$
and noticing that $\mathbf{D}\mathbf{e}^{c}$ is the torsion 2-form,
$\mathbf{T}^{c}$
\begin{eqnarray*}
0 & = & \intop\frac{\alpha}{\left(n-2\right)!}\delta\omega_{\quad k}^{ab}\mathbf{e}^{k}\wedge\left(2\varphi\left(\mathbf{D}\varphi\right)\wedge\mathbf{e}^{c}\wedge\cdots\wedge\mathbf{e}^{d}\varepsilon_{abc\cdots d}+\left(n-2\right)\varphi^{2}\mathbf{T}^{c}\wedge\mathbf{e}^{d}\wedge\cdots\wedge\mathbf{e}^{e}\varepsilon_{abcd\cdots e}\right)
\end{eqnarray*}
and therefore
\begin{eqnarray*}
0 & = & -\frac{2}{n-2}\frac{1}{\varphi}\mathbf{D}\varphi\wedge\mathbf{e}^{k}\wedge\mathbf{e}^{c}\wedge\cdots\wedge\mathbf{e}^{d}\varepsilon_{abc\cdots d}+\mathbf{T}^{c}\wedge\mathbf{e}^{k}\wedge\mathbf{e}^{d}\wedge\cdots\wedge\mathbf{e}^{e}\varepsilon_{abcd\cdots e}\\
 & = & -\frac{2}{n-2}\frac{1}{\varphi}D_{m}\varphi\mathbf{e}^{m}\wedge\mathbf{e}^{k}\wedge\mathbf{e}^{c}\wedge\cdots\wedge\mathbf{e}^{d}\varepsilon_{abc\cdots d}+\frac{1}{2}T_{\;\;mn}^{c}\mathbf{e}^{m}\wedge\mathbf{e}^{n}\wedge\mathbf{e}^{k}\wedge\mathbf{e}^{d}\wedge\cdots\wedge\mathbf{e}^{e}\varepsilon_{abcd\cdots e}\\
 & = & \left(\frac{2}{n-2}\frac{1}{\varphi}D_{m}\varphi\varepsilon^{mkc\cdots d}\varepsilon_{abc\cdots d}-\frac{1}{2}T_{\;\;mn}^{c}\varepsilon^{mnkd\cdots e}\varepsilon_{abcd\cdots e}\right)\boldsymbol{\Phi}
\end{eqnarray*}
Taking the dual eliminates the volume form. Then, resolving the pairs
of Levi-Civita tensors,
\begin{eqnarray*}
0 & = & -\frac{2}{n-2}\frac{1}{\varphi}D_{m}\varphi\left(n-2\right)!\left(\delta_{a}^{m}\delta_{b}^{k}-\delta_{a}^{k}\delta_{b}^{m}\right)\\
 &  & +\frac{1}{2}\left(n-3\right)!T_{\;\;mn}^{c}\left(\delta_{a}^{m}\left(\delta_{b}^{n}\delta_{c}^{k}-\delta_{b}^{k}\delta_{c}^{n}\right)+\delta_{a}^{n}\left(\delta_{b}^{k}\delta_{c}^{m}-\delta_{b}^{m}\delta_{c}^{k}\right)+\delta_{a}^{k}\left(\delta_{b}^{m}\delta_{c}^{n}-\delta_{b}^{n}\delta_{c}^{m}\right)\right)\\
0 & = & \frac{1}{2}\left(T_{\;\;ab}^{k}-\delta_{b}^{k}T_{\;\;an}^{n}+T_{\;\;ma}^{m}\delta_{b}^{k}-T_{\;\;ba}^{k}+2\delta_{a}^{k}T_{\;\;bc}^{c}\right)-\frac{2}{\varphi}\left(\delta_{b}^{k}D_{a}\varphi-\delta_{a}^{k}D_{b}\varphi\right)\\
 & = & T_{\;\;ab}^{k}+\delta_{b}^{k}T_{\;\;ma}^{m}-\delta_{a}^{k}T_{\;\;cb}^{c}-\frac{2}{\varphi}\left(\delta_{b}^{k}D_{a}\varphi-\delta_{a}^{k}D_{b}\varphi\right)
\end{eqnarray*}
The trace gives
\begin{eqnarray*}
0 & = & \left(n-2\right)T_{\;\;ma}^{m}-\frac{2}{\varphi}\left(n-1\right)D_{a}\varphi\\
T_{\;\;ma}^{m} & = & \frac{2}{\varphi}\frac{n-1}{n-2}D_{a}\varphi
\end{eqnarray*}
so that
\begin{eqnarray*}
T_{\;\;ab}^{k} & = & \delta_{a}^{k}T_{\;\;cb}^{c}-\delta_{b}^{k}T_{\;\;ma}^{m}+\frac{2}{\varphi}\left(\delta_{b}^{k}D_{a}\varphi-\delta_{a}^{k}D_{b}\varphi\right)\\
 & = & \frac{2}{\varphi}\left(\frac{n-1}{n-2}\delta_{a}^{k}D_{b}\varphi-\frac{n-1}{n-2}\delta_{b}^{k}D_{a}\varphi+\delta_{b}^{k}D_{a}\varphi-\delta_{a}^{k}D_{b}\varphi\right)\\
 & = & \frac{2}{\varphi}\left(\left(\frac{n-1}{n-2}-1\right)\delta_{a}^{k}D_{b}\varphi-\left(\frac{n-1}{n-2}-1\right)\delta_{b}^{k}D_{a}\varphi\right)
\end{eqnarray*}
and we arrive at a solution for the torsion,
\begin{equation}
T_{\;\;ab}^{c}=\frac{1}{n-2}\frac{2}{\varphi}\left(\delta_{a}^{c}D_{b}\varphi-\delta_{b}^{c}D_{a}\varphi\right)\label{Solution for torsion}
\end{equation}

\subsection{Collected field equations}

Collecting the variational field equations, eqs.(\ref{Einstein equation with sources}-\ref{Solution for torsion}),
and restoring the orthonormal basis,
\begin{eqnarray}
\mathfrak{R}_{ab}-\frac{1}{2}\eta_{ab}\mathfrak{R} & = & \frac{1}{\alpha}\left[-\frac{1}{\varphi^{2}}D_{a}\varphi D_{b}\varphi+\frac{1}{2}\eta_{ab}\left(\frac{1}{\varphi^{2}}D^{c}\varphi D_{c}\varphi+m^{2}+\beta\varphi^{\frac{4}{n-2}}\right)\right]\label{Einstein equation}\\
T_{\;\;ab}^{c} & = & \frac{2}{n-2}\frac{1}{\varphi}\left(\delta_{a}^{c}D_{b}\varphi-\delta_{b}^{c}D_{a}\varphi\right)\label{Torsion field equation}\\
D_{a}\varphi & = & 0\label{Constancy of phi}\\
D^{a}D_{a}\varphi-m^{2}\varphi-\alpha\mathfrak{R}\varphi-\frac{n}{n-2}\beta\varphi^{\frac{n+2}{n-2}} & = & 0\label{Wave equation}
\end{eqnarray}

It is interesting to note that the form of the torsion in eq.(\ref{Torsion field equation})
is that of Einstein's lambda transformation, which is from the most
general transformation of the connection leaving the curvature unchanged.
However, substituting eq.(\ref{Constancy of phi}) into eq.(\ref{Torsion field equation}),
we see that the torsion vanishes,
\begin{eqnarray*}
T_{\;\;ab}^{c} & = & 0
\end{eqnarray*}
Again using eq.(\ref{Constancy of phi}) in eq.(\ref{Wave equation})
get a relationship between the mass, the scalar field and the scalar
curvature,
\begin{eqnarray*}
\mathfrak{R} & = & -\frac{m^{2}}{\alpha}-\frac{n}{n-2}\frac{\beta}{\alpha}\varphi^{\frac{4}{n-2}}
\end{eqnarray*}
A similar but different relation arises when we look at the trace
of eq.(\ref{Einstein equation}) with $\varphi_{a}=0$,
\begin{eqnarray*}
\mathfrak{R}_{ab}-\frac{1}{2}\mathfrak{R}\eta_{ab} & = & \frac{1}{2\alpha\varphi^{2}}\eta_{ab}\left(m^{2}\varphi^{2}+\beta\varphi^{\frac{2n}{n-2}}\right)\\
\mathfrak{R} & = & -\frac{1}{\alpha}\frac{n}{n-2}\left(m^{2}+\beta\varphi^{\frac{4}{n-2}}\right)
\end{eqnarray*}
Comparing the two,
\begin{eqnarray*}
-\frac{m^{2}}{\alpha}-\frac{n}{n-2}\frac{\beta}{\alpha}\varphi^{\frac{4}{n-2}} & = & -\frac{1}{\alpha}\frac{n}{n-2}m^{2}-\frac{n}{n-2}\frac{\beta}{\alpha}\varphi^{\frac{4}{n-2}}\\
\frac{2}{n-2}m^{2} & = & 0
\end{eqnarray*}
so the mass of $\varphi$ must vanish in any dimension. However, there
is no constraint on $\frac{\beta}{\alpha}\varphi^{\frac{4}{n-2}}$,
which in the Riemannian gauge leads to the usual Einstein equation
with cosmological constant. 

As note following eq.(\ref{Solution for Weyl vector}), $\mathbf{D}\varphi=0$
requires that there exist a local gauge in which the Weyl vector vanishes.
In this gauge, the Weyl-Riemann tensor reduces to the usual Riemann
curvature,
\begin{eqnarray*}
\left.\boldsymbol{\mathfrak{R}}_{\;b}^{a}\right|_{W_{a}=0,T_{\;bc}^{a}=0} & = & \mathbf{R}_{\;b}^{a}
\end{eqnarray*}
and the Einstein tensor takes the usual Riemannian form. The system
has reduced to exactly the vacuum Einstein equation with cosmological
constant $\Lambda$ in a Riemannian geometry,
\begin{eqnarray*}
R_{ab}-\frac{1}{2}\eta_{ab}R-\eta_{ab}\Lambda & = & 0\\
T_{\;\;ab}^{c} & = & 0\\
\varphi & = & \varphi_{0}\\
W_{a} & = & 0
\end{eqnarray*}
where
\begin{eqnarray*}
\Lambda & \equiv & \frac{\beta}{2\alpha}\varphi_{0}^{\frac{4}{n-2}}
\end{eqnarray*}
the only difference being that in this formulation we retain local
scale invariance with an integrable Weyl vector. In a general gauge
$\phi$, the solution takes the form:
\begin{eqnarray*}
\mathfrak{R}_{ab}-\frac{1}{2}\mathfrak{R}\eta_{ab} & = & 0\\
T_{\;\;bc}^{a} & = & 0\\
W_{a} & = & \partial_{a}\phi\\
\varphi & = & \varphi_{0}e^{-\frac{n-2}{2}\phi}
\end{eqnarray*}
where this form of the scalar field $\varphi$ insures that $D_{a}\varphi=0$.
The form in a general gauge describes exactly the same physical situation,
but in a set of units which may vary from place to place.

\section{Conclusion}

We developed the properties of Weyl geometry using the Cartan formalism
for gauge theories, including enough details of the calculations to
illustrate the techniques and show the advantages of the Cartan approach.
Beginning with a review of the conformal properties of Riemannian
spacetimes, we present an efficient form of the decomposition of the
Riemann curvature into trace and traceless parts. This allows an easy
proof that the Weyl curvature tensor is the conformally invariant
part of the Riemann curvature, and shows the explicit change in the
Ricci and Schouten tensors produced by a conformal transformation.

By writing the change in the Schouten tensor as a system of differential
equations for the conformal factor, we reproduce the well-known condition
for the existence of a conformal transformation to a Ricci-flat spacetime
as a pair of integrability conditions. Continuing with the streamlined
approach, we generalize this condition to a derivation of the condition
for the existence of a conformal transformation to a spacetime satisfying
the Einstein equation with matter sources. The inclusion of matter
sources is a new result. 

Next, enlarging the symmetry from Poincaré to Weyl, we develop the
Cartan structure equations of Cartan-Weyl geometry without assuming
vanishing torsion. We find the form of the curvature tensor and its
relationship to the Riemann curvature of the corresponding Riemannian
geometry, and show that the spin connection reproduces the expected
coordinate form of Weyl connection plus contorsion tensor. We then
look at possible gravity theories. We use the Gauss-Bonnet form of
the Euler character to write the general form of quadratic-curvature
action in terms of the Ricci tensor and scalar, then turn to a detailed
description of a modified form of the Dirac scalar-tensor action.
Our approacg differs from that of either Dirac or Brans-Dicke in three
ways: we allow nonvanishing torsion, we vary the solder form and spin
connection independently, and we work in arbitrary dimension. We find
that the torsion and gradient of the scalar field must both vanish,
exactly reducing the system to locally scale-covariant general relativity
with cosmological constant.

\part*{Appendix: Bianchi identities}

The Cartan structure equations,
\begin{eqnarray*}
\mathbf{d}\boldsymbol{\omega}_{\;b}^{a} & = & \boldsymbol{\omega}_{\;b}^{c}\wedge\boldsymbol{\omega}_{\;c}^{a}+\boldsymbol{\mathfrak{R}}_{\;b}^{a}\\
\mathbf{d}\mathbf{e}^{a} & = & \mathbf{e}^{b}\wedge\boldsymbol{\omega}_{\;b}^{a}+\boldsymbol{\omega}\wedge\mathbf{e}^{a}+\mathbf{T}^{a}\\
\mathbf{d}\boldsymbol{\omega} & = & \boldsymbol{\Omega}\\
W_{\mu,\nu}\mathbf{d}x^{\nu}\wedge\mathbf{d}x^{\mu} & = & \frac{1}{2}\Omega_{\mu\nu}\mathbf{d}x^{\mu}\wedge\mathbf{d}x^{\nu}\\
W_{\mu,\nu}-W_{\nu,\mu} & = & \Omega_{\mu\nu}
\end{eqnarray*}
have integrability conditions, which for gravity theories are called
Bianchi identities. These follow from the Poincaré lemma, $\mathbf{d}^{2}\equiv0$:
\begin{eqnarray*}
\mathbf{d}\boldsymbol{\mathfrak{R}}_{\;b}^{a} & = & \mathbf{d}^{2}\boldsymbol{\omega}_{\;b}^{a}-\mathbf{d}\boldsymbol{\omega}_{\;b}^{c}\wedge\boldsymbol{\omega}_{\;c}^{a}+\boldsymbol{\omega}_{\;b}^{c}\wedge\mathbf{d}\boldsymbol{\omega}_{\;c}^{a}\\
0 & = & \mathbf{d}\boldsymbol{\mathfrak{R}}_{\;b}^{a}+\left(\boldsymbol{\mathfrak{R}}_{\;b}^{c}+\boldsymbol{\omega}_{\;b}^{e}\wedge\boldsymbol{\omega}_{\;e}^{c}\right)\wedge\boldsymbol{\omega}_{\;c}^{a}-\boldsymbol{\omega}_{\;b}^{c}\wedge\left(\boldsymbol{\mathfrak{R}}_{\;c}^{a}+\boldsymbol{\omega}_{\;c}^{e}\wedge\boldsymbol{\omega}_{\;e}^{a}\right)\\
 & = & \mathbf{d}\boldsymbol{\mathfrak{R}}_{\;b}^{a}+\boldsymbol{\mathfrak{R}}_{\;b}^{c}\wedge\boldsymbol{\omega}_{\;c}^{a}-\boldsymbol{\mathfrak{R}}_{\;c}^{a}\wedge\boldsymbol{\omega}_{\;b}^{c}\\
\mathbf{D}\boldsymbol{\mathfrak{R}}_{\;b}^{a} & = & 0\\
\mathbf{d}^{2}\mathbf{e}^{a} & = & \mathbf{d}\mathbf{e}^{b}\wedge\boldsymbol{\omega}_{\;b}^{a}-\mathbf{e}^{b}\wedge\mathbf{d}\boldsymbol{\omega}_{\;b}^{a}+\mathbf{d}\boldsymbol{\omega}\wedge\mathbf{e}^{a}-\boldsymbol{\omega}\wedge\mathbf{d}\mathbf{e}^{a}+\mathbf{d}\mathbf{T}^{a}\\
0 & = & \left(\mathbf{e}^{c}\wedge\boldsymbol{\omega}_{\;c}^{b}+\boldsymbol{\omega}\wedge\mathbf{e}^{b}+\mathbf{T}^{b}\right)\wedge\boldsymbol{\omega}_{\;b}^{a}-\mathbf{e}^{b}\wedge\left(\boldsymbol{\omega}_{\;b}^{c}\wedge\boldsymbol{\omega}_{\;c}^{a}+\boldsymbol{\mathfrak{R}}_{\;b}^{a}\right)+\boldsymbol{\Omega}\wedge\mathbf{e}^{a}-\boldsymbol{\omega}\wedge\left(\mathbf{e}^{b}\wedge\boldsymbol{\omega}_{\;b}^{a}+\boldsymbol{\omega}\wedge\mathbf{e}^{a}+\mathbf{T}^{a}\right)+\mathbf{d}\mathbf{T}^{a}\\
0 & = & -\mathbf{e}^{b}\wedge\boldsymbol{\mathfrak{R}}_{\;b}^{a}+\boldsymbol{\Omega}\wedge\mathbf{e}^{a}+\mathbf{d}\mathbf{T}^{a}+\mathbf{T}^{b}\wedge\boldsymbol{\omega}_{\;b}^{a}-\boldsymbol{\omega}\wedge\mathbf{T}^{a}\\
\mathbf{D}\mathbf{T}^{a} & = & \mathbf{e}^{b}\wedge\boldsymbol{\mathfrak{R}}_{\;b}^{a}-\boldsymbol{\Omega}\wedge\mathbf{e}^{a}\\
\mathbf{d}\boldsymbol{\Omega} & = & 0
\end{eqnarray*}
Summary:
\begin{eqnarray*}
\mathbf{D}\boldsymbol{\mathfrak{R}}_{\;b}^{a} & \equiv & \mathbf{d}\boldsymbol{\mathfrak{R}}_{\;b}^{a}+\boldsymbol{\mathfrak{R}}_{\;b}^{c}\wedge\boldsymbol{\omega}_{\;c}^{a}-\boldsymbol{\mathfrak{R}}_{\;c}^{a}\wedge\boldsymbol{\omega}_{\;b}^{c}\equiv0\\
\mathbf{D}\mathbf{T}^{a} & = & \mathbf{e}^{b}\wedge\boldsymbol{\mathfrak{R}}_{\;b}^{a}-\boldsymbol{\Omega}\wedge\mathbf{e}^{a}\\
\mathbf{D}\boldsymbol{\Omega} & = & 0
\end{eqnarray*}
where
\begin{eqnarray*}
\mathbf{D}\boldsymbol{\mathfrak{R}}_{\;b}^{a} & \equiv & \mathbf{d}\boldsymbol{\mathfrak{R}}_{\;b}^{a}+\boldsymbol{\mathfrak{R}}_{\;b}^{c}\wedge\boldsymbol{\omega}_{\;c}^{a}-\boldsymbol{\mathfrak{R}}_{\;c}^{a}\wedge\boldsymbol{\omega}_{\;b}^{c}\\
\mathbf{D}\mathbf{T}^{a} & \equiv & \mathbf{d}\mathbf{T}^{a}+\mathbf{T}^{b}\wedge\boldsymbol{\omega}_{\;b}^{a}-\boldsymbol{\omega}\wedge\mathbf{T}^{a}\\
\mathbf{D}\boldsymbol{\Omega} & \equiv & \mathbf{d}\boldsymbol{\Omega}
\end{eqnarray*}

When the torsion vanishes, we have an algebraic identity,
\begin{eqnarray*}
\mathbf{e}^{b}\boldsymbol{\mathfrak{R}}_{\;b}^{a} & = & \boldsymbol{\Omega}\wedge\mathbf{e}^{a}\\
\mathfrak{R}_{\;\left[bcd\right]}^{a} & = & \delta_{[b}^{a}\Omega_{cd]}\\
\mathfrak{R}_{\;bcd}^{a}+\mathfrak{R}_{\;cdb}^{a}+\mathfrak{R}_{\;dbc}^{a} & = & \delta_{b}^{a}\Omega_{cd}+\delta_{c}^{a}\Omega_{db}+\delta_{d}^{a}\Omega_{bc}\\
\mathfrak{R}_{bd}-\mathfrak{R}_{bd} & = & -\left(n-2\right)\Omega_{bd}
\end{eqnarray*}

\end{document}